\documentclass[a4paper,fleqn,usenatbib,twocolumn]{mnras}

\usepackage{newtxtext,newtxmath}

\usepackage[T1]{fontenc}
\usepackage{ae,aecompl}

\usepackage{graphicx}	% Including figure files
\usepackage{amsmath}	% Advanced maths commands
\usepackage{amssymb}	% Extra maths symbols
\usepackage{placeins}
\usepackage[usenames,dvipsnames]{color} % Allows for comments

\def\code#1{\texttt{#1}}

\newcommand{\brvs}{Br\"unt-V\"ais\"al\"a}
\newcommand{\alf}{Alfv\'{e}n}
\title[Convective Differential Rotation II]{Convective Differential Rotation in Stars and Planets II: Observational and Numerical Tests}

\author[Adam S. Jermyn]{
Adam S. Jermyn$^{1,2}$\thanks{E-mail: adamjermyn@gmail.com}
Shashikumar M. Chitre,$^{2,3}$
Pierre Lesaffre$^{4}$
and
Christopher A. Tout$^{2}$
\\
$^{1}$Center for Computational Astrophysics, Flatiron Institute, New York, New York, 10010, USA\\
$^{2}$Institute of Astronomy, University of Cambridge, Madingley Road, Cambridge CB3 0HA, UK\\
$^{3}$Centre for Basic Sciences, University of Mumbai, India\\
$^{4}$\'{E}cole Normale Sup\'{e}rieure 24 rue Lhomond, 75231 Paris, France
}

\date{Accepted XXX. Received YYY; in original form ZZZ}

\pubyear{2020}

\begin{document}
\label{firstpage}
\pagerange{\pageref{firstpage}--\pageref{lastpage}}
\maketitle

\begin{abstract}
Differential rotation is central to a great many mysteries in stars and planets.
In Part I we predicted the order of magnitude and scaling of the differential rotation in both hydrodynamic and magnetohydrodynamic convection zones.
Our results apply to both slowly- and rapidly-rotating systems, and provide a general picture of differential rotation in stars and fluid planets.
We further calculated the scalings of the meridional circulation, entropy gradient and baroclinicity.
In this companion paper we compare these predictions with a variety of observations and numerical simulations.
With a few exceptions we find that these are consistent in both the slowly-rotating and rapidly-rotating limits.
Our results help to localize core-envelope shear in red~giant stars, suggest a rotation-dependent frequency shift in the internal gravity waves of massive stars and potentially explain observed deviations from von Zeipel's gravity darkening in late-type stars.
\end{abstract}


%%%%%%%%%%% jump to intro
%%%---------- open: intro

\section{Introduction}

Driven largely by high-cadence precision photometry from the \emph{CoRoT}~\citep{1999ASPC..173..357R}, \emph{Kepler}~\citep{Gilliland_2010} and \emph{TESS}~\citep{2015JATIS...1a4003R} missions, the ability of observations to reveal stellar rotation has increased dramatically in recent years.
Asteroseismology now enables strong constraints to be placed on radial rotation profiles~\citep{doi:10.1146/annurev-astro-091918-104359}, revealing large differential rotation in both red giant~\citep{2012Natur.481...55B} and Sun-like stars~\citep{2019A&A...626A.121O}.
Similarly, star spot timing has provided measurements of latitudinal shear at the surfaces of stars~\citep{1996ApJ...466..384D,2015A&A...583A..65R,2017AJ....154..250L}.

Spectral deconvolution~\citep{1997MNRAS.291....1D} and spectropolarimetry~\citep{2003A&A...412..813R} also provide a handle on latitudinal differential rotation and, because these do not require the presence of spots, they can be applied to a wider range of stars.
Importantly, the use of spectra rather than spots also removes uncertainty in the spot latitude.

The breadth of these observations of other stars complements the depth of those of the Sun, which reveal its detailed rotation profile~\citep{1991LNP...388...61T,1998ApJ...505..390S}.
In addition, some compact objects now provide strong constraints on the rotation profiles of their progenitors.
For instance, gravitational waves allow the spins of merging black hole binaries to be measured~\citep{2016MNRAS.462..844K,2018arXiv181112940T,2019PhRvD.100b3007Z}.
Assuming no significant spin changes owing to accretion or supernovae, such measurements then indicate the angular momenta of the cores of the progenitor stars.
Similarly, photometry of white dwarfs provides rotation rates and hence constrains the spins of cores of lower-mass stars~\citep{2017ApJS..232...23H}.

The growing diversity and depth of observations makes the distribution of angular momentum a key theoretical question which has driven the development of mean field turbulence theories~\citep{1995A&A...299..446K}, thermal wind balance arguments~\citep{2010A&A...510A..33B,2012MNRAS.420.2457B} and numerical simulations~\citep{ASNA:ASNA201012345,doi:10.1146/annurev.fluid.010908.165215}.
Importantly, rotation and differential rotation play an active role in the structure and evolution of stars.
For instance by inducing mixing~\citep{1929MNRAS..90...54E,1998ASPC..131...85M,1992A&A...253..173C} and generating magnetic fields~\citep{2002A&A...381..923S} and activity~\citep{2016Natur.535..526W}.

Previously~\citep[][hereinafter Paper~I]{part1} we provided predictions for the magnitude of differential rotation in the convection zones of stars and gaseous planets.
These predictions are order-of-magnitude scaling relations based on considerations of the asymptotic scalings of different physical processes.
Along the way we also predicted the scaling of other quantities such as the baroclinicity and meridional circulation.
In this companion paper we show that our predictions are generally in good agreemnt with a variety of different observations and numerical simulations.
The greatest disagreements are with the simulations, which arise primarily when these are highly diffusive, highlighting the importance of developed turbulence in angular momentum transport.

In the next section we define our notation and review the key assumptions and results of Paper~I.
We then proceed to compare our predictions to observations of both radial and latitudinal shear in low mass, solar-type and red giant stars, as well as {\emph{Juno}} measurements of latitudinal shear in Jupiter (section~\ref{sec:obs}).
We also compare our predictions of baroclinicity with measurements of the solar latitudinal temperature gradient.
Following this we turn to differential rotation in hydrodynamic and MHD simulations in section~\ref{sec:sims}, where we also examine related quantities such as the baroclinicity, convection speed and, for MHD simulations, the magnetic energy density.
In section~\ref{sec:future} we describe further tests which could be done given more observations and numerical simulations.
We conclude with a discussion of the results and their astrophysical implications in section~\ref{sec:discussion}.%%%---------- close: intro

%%%%%%%%%%% jump to definitions
%%%---------- open: definitions
\section{Overview}
\label{sec:definitions}

\begin{figure}
\centering
\includegraphics[width=0.4\textwidth]{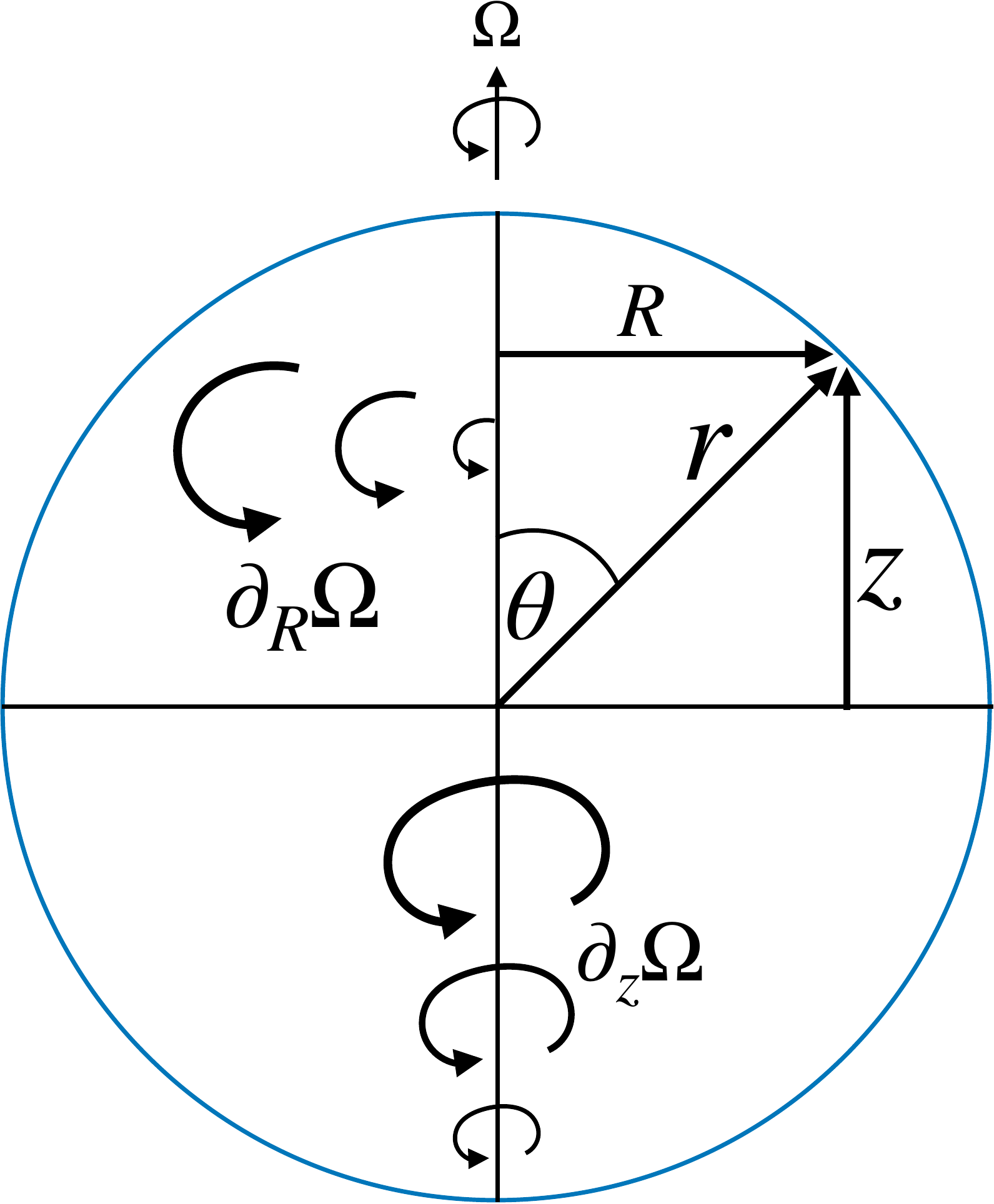}
\caption{The rotation, coordinate system and differential rotation are shown schematically: (top) the mean angular velocity $\Omega$; (upper-right) the cylindrical radius $R$, vertical direction along the rotation axis $z$, spherical radius $r$ and polar angle $\theta$; (upper-left) an example of cylindrical radial differential rotation ($\partial_R \Omega$); (lower) an example of cylindrical vertical differential rotation ($\partial_z\Omega$).}
\label{fig:schema3}
\end{figure}

\begin{figure}
\centering
\includegraphics[width=0.4\textwidth]{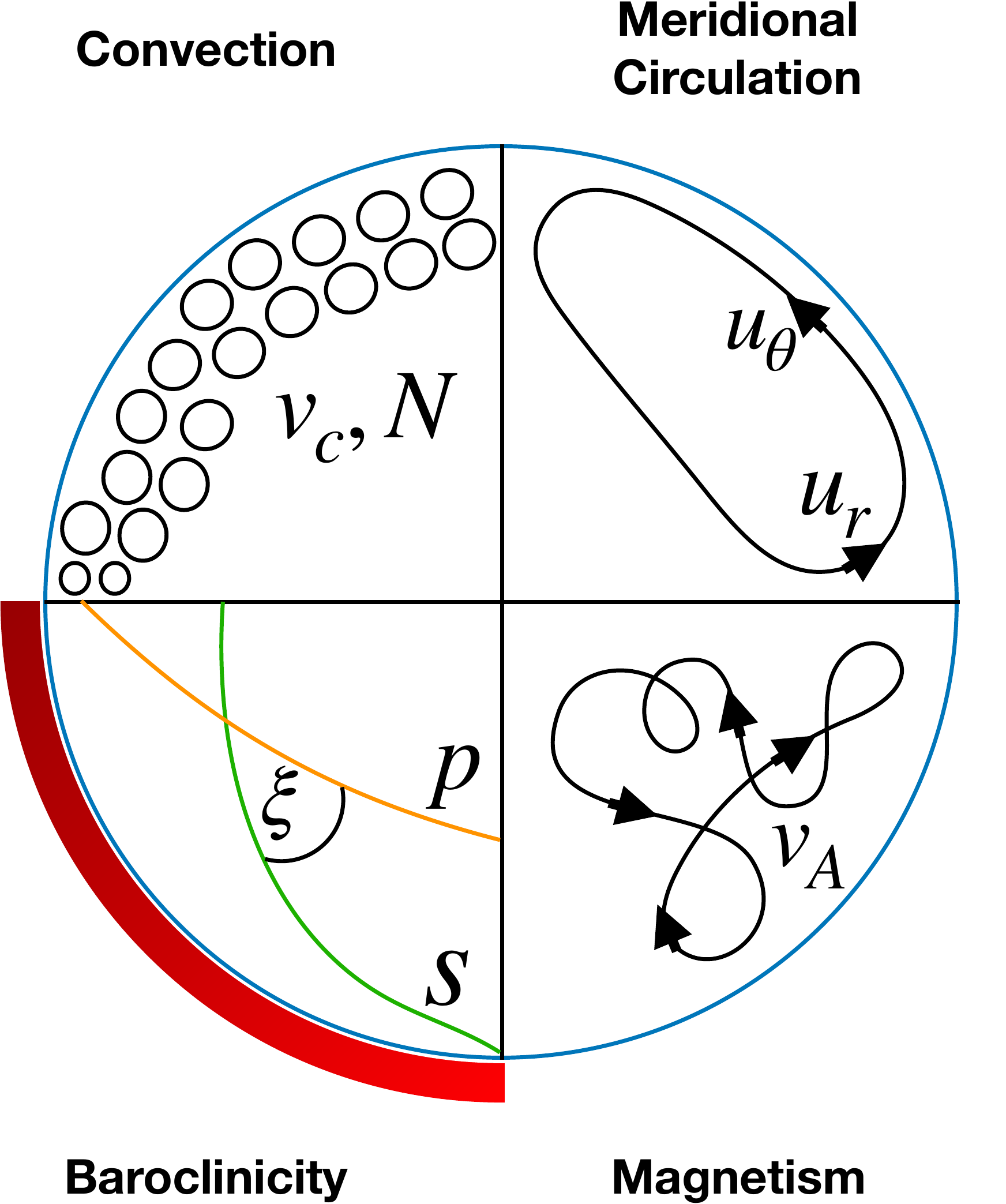}
\caption{Key concepts in our theory are shown schematically: (upper-right) the meridional circulation components $u_r$ and $u_\theta$; (upper-left) turbulent eddies move at the convection speed $\varv_c$ and with the \brvs\ frequency $N$; (lower-left) surfaces of constant pressure and entropy meet at an angle of approximately the baroclinicity $\xi$, resulting in surface temperature variations; (lower-right) a magnetic field with \alf\ velocity $\varv_{\rm A}$.}
\label{fig:schema4}
\end{figure}

Rotation breaks spherical symmetry.
This means that rotating stars may be very different from non-rotating stars.
In Paper I we studied the response of convection zones to both slow and rapid rotation.
We focused on systems with \emph{efficient} convection zones, meaning those in which most of the heat is transported by advection and very little by diffusion.
In such systems we expect the microscopic thermal diffusivity, composition diffusivity and viscosity to all be irrelevant, which allows us to specify the problem fully by giving just the geometry of the convection zone, the total angular momentum and the profile of the \brvs\ frequency
\begin{align}
	N^2 &= -\frac{\gamma-1}{\gamma} \boldsymbol{g}\cdot\nabla s,
	\label{eq:N2}
\end{align}
where $\boldsymbol{g}$ is the acceleration due to gravity, which need not be radial, $s$ is the dimensionless entropy\footnote{See Appendix~B of Paper~I.}
\begin{align}
	s = \frac{1}{\gamma-1}\left(\ln P - \gamma \ln \rho\right),
	\label{eq:s}
\end{align}
where $P$ is the pressure, $\rho$ is the density, and $\gamma$ is the adiabatic index.

Unfortunately, turbulence is difficult to analyze.
Because of this we just attempted to understand the orders of magnitude and scaling relations involved in rotating convection.
With this, the problem simplifies to one specified just by the geometry, the ratio of the angular velocity $\Omega$ to the magnitude of the \brvs\ frequency $|N|$, and the ratio of the pressure scale height
\begin{align}
	h \equiv |\nabla\ln P|^{-1}
	\label{eq:h}
\end{align}
to the spherical radius $r$.
In the same spirit we further assume that the precise geometry does not matter, and treat spherical shells and fully spherical convection zones alike.

Our analysis began in the non-rotating limit, which on average must be spherically symmetric.
In that limit the convection zone is parameterized by the scale height $h$ and the \brvs\ frequency $|N|$.
From these we then expect to determine the convection speed $\varv_{\rm c}$, which is the root-mean-square of the difference between the velocity of a fluid element and its mean velocity, as well as the \alf\ velocity
\begin{align}
	\varv_{\rm A} \equiv \frac{B}{\sqrt{4\pi \rho}},
\end{align}
where $B$ is the root-mean-squared magnetic field strength.
These quantities are shown in the upper-left and lower-right panels of Fig.~\ref{fig:schema4}.
In ordinary mixing length theory for efficient convection $\varv_{\rm c} \approx h |N|$, and $\varv_{\rm A}$ is usually expected to be of the same order~\citep{2019ApJ...883..106C}, so we assume both of these hold in the non-rotating limit.

We next turned to the limit of slow rotation ($\Omega \ll |N|$).
In this limit rotation breaks spherical symmetry.
This allows there to be steady meridional circulation currents, differential rotation, and misaligned entropy and pressure gradients, shown in Fig.~\ref{fig:schema3} and the upper-right and lower-left panels of Fig.~\ref{fig:schema4}.
For convenience in working with these quantities we defined the baroclinicity
\begin{align}
	\xi \equiv \frac{\boldsymbol{e}_\phi \cdot \left(\nabla \ln P \times \nabla s\right)}{|\nabla \ln P| |\nabla s|},
	\label{eq:xi}
\end{align}
which is directly proportional to the thermal wind term in the vorticity equation (Part~1 equation~3) and, when small, is approximately the angle between the pressure and entropy gradients.

To determine how each of these quantities scales with $\Omega$ we used symmetry arguments to constrain the possible scalings of the different components of the turbulent Reynolds stress.
So, for instance, because mapping $\Omega \rightarrow -\Omega$ is equivalent to mapping the azimuthal coordinate $\phi \rightarrow -\phi$, we know that the Reynolds stress component $\mathbf{T}_{r\theta}$ must be an even function of $\Omega$ because it is even under $\phi \rightarrow -\phi$.
If the stress is analytic in a region around $\Omega=0$ this means that $\mathbf{T}_{r\theta}$ scales at least as fast as $\Omega^2$.
We then assumed that each quantity scales at the lowest allowed order and obtained scalings for each of the quantities shown in Figs.~\ref{fig:schema3} and~\ref{fig:schema4}, including the shear and baroclinicity.
These scalings suggest, that in steady state, the thermal wind term, mechanical forcing by turbulent stress, and losses owing to turbulent viscosity are all of the same order and serve to set the shear.

In the limit of rapid rotation we then assumed that there are no remaining symmetries, so that all components of the stress and turbulent diffusivity and so on are of the same order.
We supplemented this with the scaling laws of~\citet{1979GApFD..12..139S} for the convection speed and~\citet{doi:10.1111/j.1365-246X.2006.03009.x} for the \alf\ velocity.
Importantly these scalings imply that the \brvs\ frequency depends on the rotation rate in this limit, so we denote the non-rotating frequency as $|N|_0$.

We were then able to determine the lowest possible order of each quantity in $\Omega^{-1}$, and assumed that each declines as slowly as possible given these constraints.
The resulting scaling laws indicate that the Taylor-Proudman term $\frac{\partial \Omega^2}{\partial z}$ balances azimuthal turbulent stresses and thereby sets the differential rotation, and that advection of angular momentum balances meridional turbulent stresses to set the meridional circulation.

To summarize, our predictions were derived under a few assumptions.
\begin{enumerate}
	\item Dimensionless factors arising from geometry are of order unity unless symmetries require them to be otherwise.
	\item All external perturbing forces, such as tides or external heating, are negligible in the regions of interest.
	\item The material is non-degenerate, compressible and not radiation-dominated.
	\item All microscopic (i.e. non-turbulent) diffusivities are negligible, such that
	\begin{enumerate}
	\item convection is efficient, so the gas is nearly isentropic,
	\item the Reynolds and Rayleigh numbers are much larger than critical, and
	\item magnetohydrodynamical processes are ideal.
	\end{enumerate}
	\item The system is axisymmetric in a time-averaged sense.
	\item Convection is subsonic.
	\item The system is chemically homogeneous.
	\item The pressure scale height $h$ is less than or of the same order as the radius $r$.
\end{enumerate}
Most of these predictions hold in all of the systems with which we test our predictions.
The two exceptions are the assumption that convection is subsonic, which is marginal in systems with shallow surface convection zones, and that $h la r$, which fails to hold in near the centres of core-convecting stars.
However, when these fail to hold they usually do so over a small volume, so we do not expect either of them to prevent us from comparing our predictions with observations or numerical simulations.
%%%---------- close: definitions

%%%%%%%%%%% jump to observations
%%%---------- open: observations

\section{Observational Tests}
\label{sec:obs}

\begin{table*}
\caption{The scalings of the differential rotation, meridional circulation, baroclinicity, \brvs\ frequency, convective velocity, and the ratio of magnetic to kinetic energy are given for the three regimes of interest in terms of the non-rotating \brvs\ frequency $|N|_0$.}
\label{tab:summary}
\begin{tabular}{lllllll}
\hline
\hline
Case & $\frac{|R\nabla \Omega|}{\Omega}$ & $\frac{|R\partial_R \Omega|}{\Omega}$ & $\frac{|R\partial_z \Omega|}{\Omega}$ & $\frac{|r\partial_r \Omega|}{\Omega}$ &$\frac{|\partial_\theta \Omega|}{\Omega}$ &\\
\hline
Slow ($\Omega \ll |N|_0$) & $1$ & $1$ & $1$ & $1$ & $1$&\\
Fast Hydro.($\Omega \gg |N|_0$) & $\left(\frac{\Omega}{|N|_0}\right)^{-3/5}$& $\left(\frac{\Omega}{|N|_0}\right)^{-3/5}$ & $\left(\frac{\Omega}{|N|_0}\right)^{-6/5}$& $\left(\frac{\Omega}{|N|_0}\right)^{-3/5}$& $\left(\frac{\Omega}{|N|_0}\right)^{-3/5}$&\\
Fast MHD ($\Omega \gg |N|_0$) &$\left(\frac{\Omega}{|N|_0}\right)^{-3/4}$&$\left(\frac{\Omega}{|N|_0}\right)^{-3/4}$& $\left(\frac{\Omega}{|N|_0}\right)^{-3/2}$ &$\left(\frac{\Omega}{|N|_0}\right)^{-3/4}$&$\left(\frac{\Omega}{|N|_0}\right)^{-3/4}$&\\
\hline
\hline
Case & $\frac{u_r}{h|N|_0}$ & $\frac{u_\theta}{h|N|_0}$ & $\xi$ & $\frac{|N|}{|N|_0}$ &$\frac{\varv_{\rm c}}{h|N|_0}$ & $\frac{\varv_{\rm A}^2}{\varv_{\rm c}^2}$\\
\hline
Slow ($\Omega \ll |N|_0$) & $\frac{h}{r}\left(\frac{\Omega}{|N|_0}\right)^{2}$ & $\left(\frac{\Omega}{|N|_0}\right)^{2}$ & $\left(\frac{\Omega}{|N|}\right)^2$ & $1$ & $1$ & $1$\\
Fast Hydro.($\Omega \gg |N|_0$) & $\frac{h}{r}\left(\frac{\Omega}{|N|_0}\right)^{-7/5}$& $\left(\frac{\Omega}{|N|_0}\right)^{-7/5}$ & $1$ & $\left(\frac{\Omega}{|N|_0}\right)^{2/5}$ & $\left(\frac{\Omega}{|N|_0}\right)^{-1/5}$ & N/A\\
Fast MHD ($\Omega \gg |N|_0$) & $\frac{h}{r}\left(\frac{\Omega}{|N|_0}\right)^{-1/2}$ & $\left(\frac{\Omega}{|N|_0}\right)^{-1/2}$& $1$ & $\left(\frac{\Omega}{|N|_0}\right)^{1/4}$ & $\left(\frac{\Omega}{|N|_0}\right)^{-1/2}$ & $\left(\frac{\Omega}{|N|_0}\right)^{3/2}$\\
\hline
\hline
\end{tabular}
\end{table*}

In Paper I we studied differential rotation in both slowly and rapidly rotating convection zones in both the hydrodynamic and magnetohydrodynamic (MHD) limits.
Our results are given in Table~\ref{tab:summary}.
In the limit of slow rotation the scaling is the same for both hydrodynamic and MHD convection so those cases are grouped together.

It is important to emphasize that we have only predicted the scalings of these various quantities but not their actual magnitudes.
We expect that in each case there are likely factors of order unity in these relations, such that for a quantity $Q$ scaling as $(\Omega/|N|_0)^\alpha$ we have
\begin{align}
	Q = \lambda_Q Q_0 \left(\frac{\Omega}{|N|_0}\right)^\beta,
\end{align}
for some $\lambda_Q$ of order unity and dimensional $Q_0$ which gives the characteristic scale in terms of $h$ and $|N|_0$.
For clarity we have non-dimensionalized all quantities appearing in Table~\ref{tab:summary}, so that in each case $Q_0 = 1$.

To compute $\Omega/|N|_0$ from observations we used given values where this quantity was provided directly, and in all other cases we first computed stellar models to match the observed characteristics of each star using the Modules for Experiments in Stellar Astrophysics
\citep[\code{MESA}][]{Paxton2011, Paxton2013, Paxton2015, Paxton2018, Paxton2019} software instrument.
We then constructed a volume-weighted average \brvs\ frequency given as
\begin{align}
	|N|_0 \approx |N|_{\rm avg} \equiv \exp\left(\frac{\int dr \frac{\ln |N|}{\max(h,r)}}{\int \frac{dr}{\max(h,r)}}\right).
	\label{eq:N_avg}
\end{align}
A geometric average was chosen because the rotation profile we predict varies as a power-law in $|N|$.
The weighting of $dr/\max(h,r)$ was chosen because the shear we calculate is per unit $\ln r$, but near the cores of stars $r < h$ and the convective cells become large relative to $r$.
In that limit $h$ becomes the relevant length-scale, so we weight by $h$ instead of $r$.

As described in Appendix~\ref{appen:data}, equation~\eqref{eq:N_avg} is the form we have used whenever a volume average of $|N|$ was needed.
In Appendix~\ref{appen:Navg} we study the sensitivity of our analysis to this averaging and find that, while there is a potentially large systematic offset in $|N|_0$ depending on the averaging used, up to a factor of $10^2$, the shape of the data are robust to different choices of averaging.
An additional factor of order $2$ uncertainty arises from uncertainties in matching stellar models to the observations, and particularly in uncertainties in the mixing length parameter.
These effects are discussed in more detail in Appendix~\ref{appen:models}.

In what follows we endeavour to test the relations in Table~\ref{tab:summary} through many different means.
We use data from different sources and so must standardize them.
Details of how the data from various sources were obtained, standardized and processed are provided in Appendix~\ref{appen:data}, and all scripts and files needed to reproduce this analysis are available on \url{https://zenodo.org/record/3992228}.

%%%%%%%%% jump to latitudinal_tests
%%%---------- open: latitudinal_tests

\subsection{Latitudinal Shear}
\label{sec:latitude}

\begin{figure}
\centering
\includegraphics[width=0.47\textwidth]{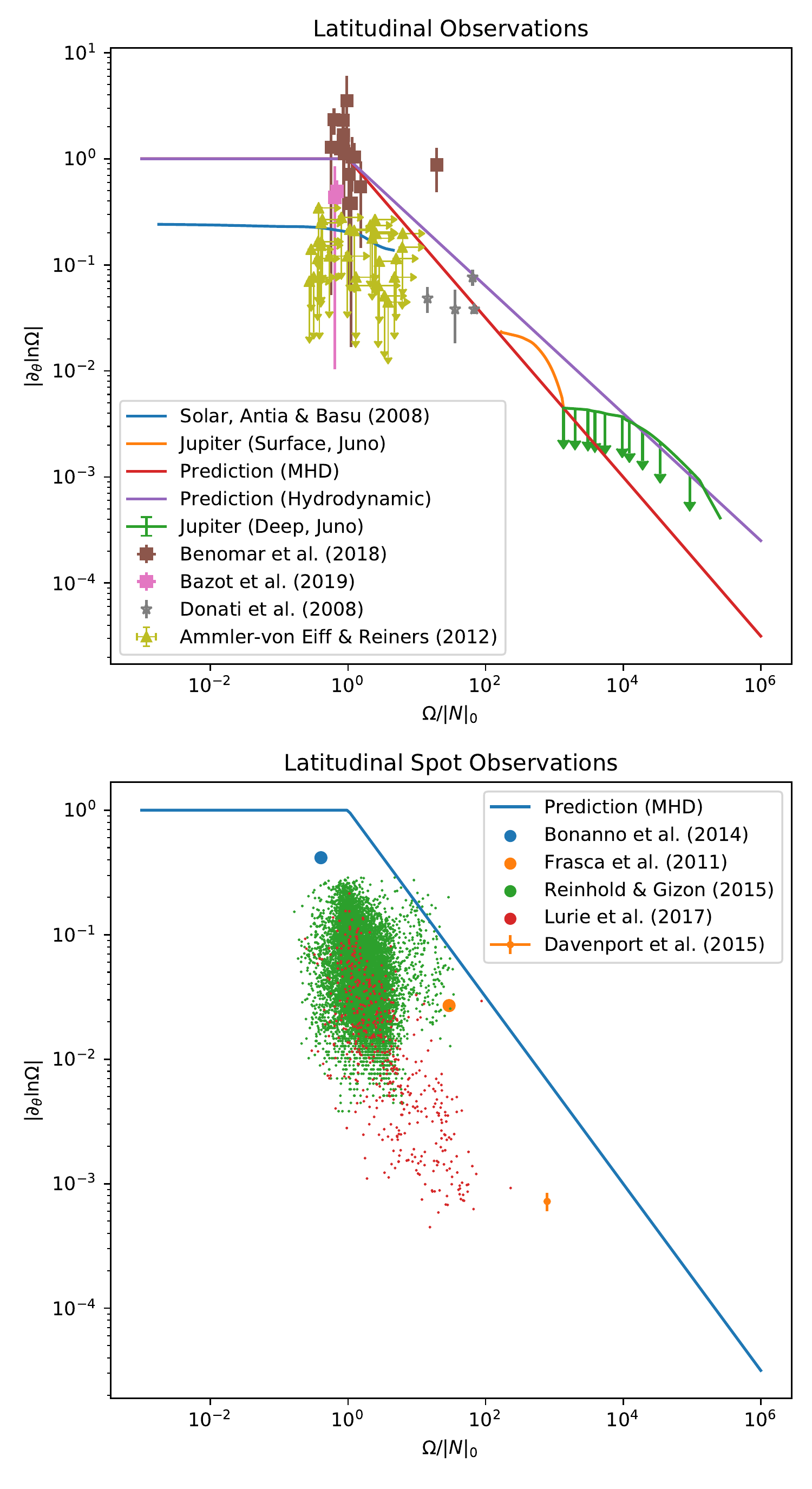}
\caption{The relative latitudinal shear $|\partial_\theta \ln\Omega|$ is shown as a function of $\Omega/|N|_0$ for observed convecting stars and Jupiter alongside our prediction, which is constant for $\Omega < |N|_0$ and scales as $(\Omega/|N|_0)^{-3/4}$ (MHD) and $(\Omega/|N|_0)^{-3/5}$ (hydrodynamic) for $\Omega > |N|_0$. Shapes indicate the origin of the data: squares are for asteroseismology, triangles are for spectroscopy, stars are for spectropolarimetry and circles are for star spot measurements. For clarity the star spot measurements are shown separately in the lower panel. Helioseismic data and constraints from {\emph{Juno}} are shown as solid lines. {\emph{Juno}} measurements of the shear at a depth of less than about $3000\rm{km}$ are shown separately from {\emph{Juno}} upper limits on that at greater depths.}
\label{fig:latitudinal}
\end{figure}

We begin by considering observations of latitudinal shear $|\partial_\theta \ln \Omega|$, shown in Fig.~\ref{fig:latitudinal} alongside our predictions for the MHD and hydrodynamic limits where relevant.
The top panel shows observations from a variety of sources including asteroseismology but excluding starspot measurements, while the bottom panel shows only starspot data.
In general these measurements should be interpreted as indicating the average shear over the surface of a star, though different techniques are sensitive to different latitudes.
For instance starspot measurements are most sensitive to shear over typical active latitudes.
While we have made an attempt to standardize these data, we urge the reader to see Appendix~\ref{appen:data} for details on how data from each source were interpreted.

With the exception of contraints from the {\emph{Juno}} mission, the top panel mostly probes the limits of slow and moderate rotation.
While there are offsets between measurements made with different techniques, there is general agreement that the shear is of order unity for slow rotators and starts to decline roughly as predicted as the rotation rate increases.
The data from~\citet{2018Sci...361.1231B} are notable outliers which lie well above measurements of objects at comparable rotation rates but otherwise the data seem generally consistent both internally and with our predictions.

The {\emph{Juno}} data provide our strongest check on the rapidly-rotating limit~\citep{2018Natur.555..223K}.
The surface data are more consistent with the hydrodynamic regime, while the upper limits from deeper regions are more consistent with the MHD regime.
This is in agreement with previous work~\citep{1987ApJ...316..836K,2008Icar..196..653L} and the finding that the transition between shallow and deep differential rotation in Jupiter is associated with a large change in electrical conductivity~\citep{2018Natur.555..227G}.
Nonetheless, even for these rapid rotation rates the difference between these scalings is only a factor of a few and this makes definitive statements about different regimes difficult.

The lower panel of Fig.~\ref{fig:latitudinal} shows starspot measurements of latitudinal shear.
The data appear consistent with our expected scaling relations.
In particular the slopes of the data both of~\citet{2015A&A...583A..65R} and~\citet{2017AJ....154..250L} roughly match our predictions, though with considerable scatter which makes it difficult to say for certain, and the three points from other sources~\citep{2011A&A...532A..81F,2014A&A...569A.113B,2015ApJ...806..212D} scale similarly.
There appears to be an overall offset of a factor of $0.1$ to $0.3$ in the magnitude of the shear between the data and our predictions, so we suggest using our scaling relations with a prefactor of $\beta_{|R\nabla\Omega|} \approx 0.2$.
This is consistent with the offset we see with the Sun, though the data of~\citet{2018Sci...361.1231B} and~\citet{2012A&A...542A.116A} suggest no such offset.
This could be due to differences in which regions of the star are probed by different techniques or it could represent systematic differences of which we are unaware in either the different measurement techniques or the objects observed.
For instance different samples may contain different fractions of fully convective stars, stars with deep convection zones, and others with shallow convection zones.
While our predictions do not depend on the depth of the convection zone that does not mean that there is no such dependence.

A further test is provided by the empirical scaling relations of~\citet{2016MNRAS.461..497B} who find that
\begin{align}
	\partial_\theta \ln \Omega \propto \Omega^{-n},
\end{align}
where $n=1.1$ for K, $0.8$ for G and $0.6$ for F stars.
Their sample is comprised mostly of stars with rotation periods of order $10\,{\rm d}$.
Except perhaps very near the surface, this is generally more rapid than the convective turnover time in these stars, so they are primarily in the rapid regime in which, for ionized systems, we predict $n=-3/4 =-0.75$, which falls in the middle of the observed range.

The trend of increasing $n$, and hence increasing relative shear, as stellar temperature rises could be due to a combination of the shallowing convection zone and an increasing \brvs\ frequency.
As this happens the mean convective motions become faster and the stars approach the slow regime, where we predict $n=0$.
So it seems likely that the trend in $n$ is a result of the transition from one regime to the other.%%%---------- close: latitudinal_tests

%%%%%%%%% jump to radial_tests
%%%---------- open: radial_tests
\subsection{Radial Shear}
\label{sec:radial}

\begin{figure}
\centering
\includegraphics[width=0.47\textwidth]{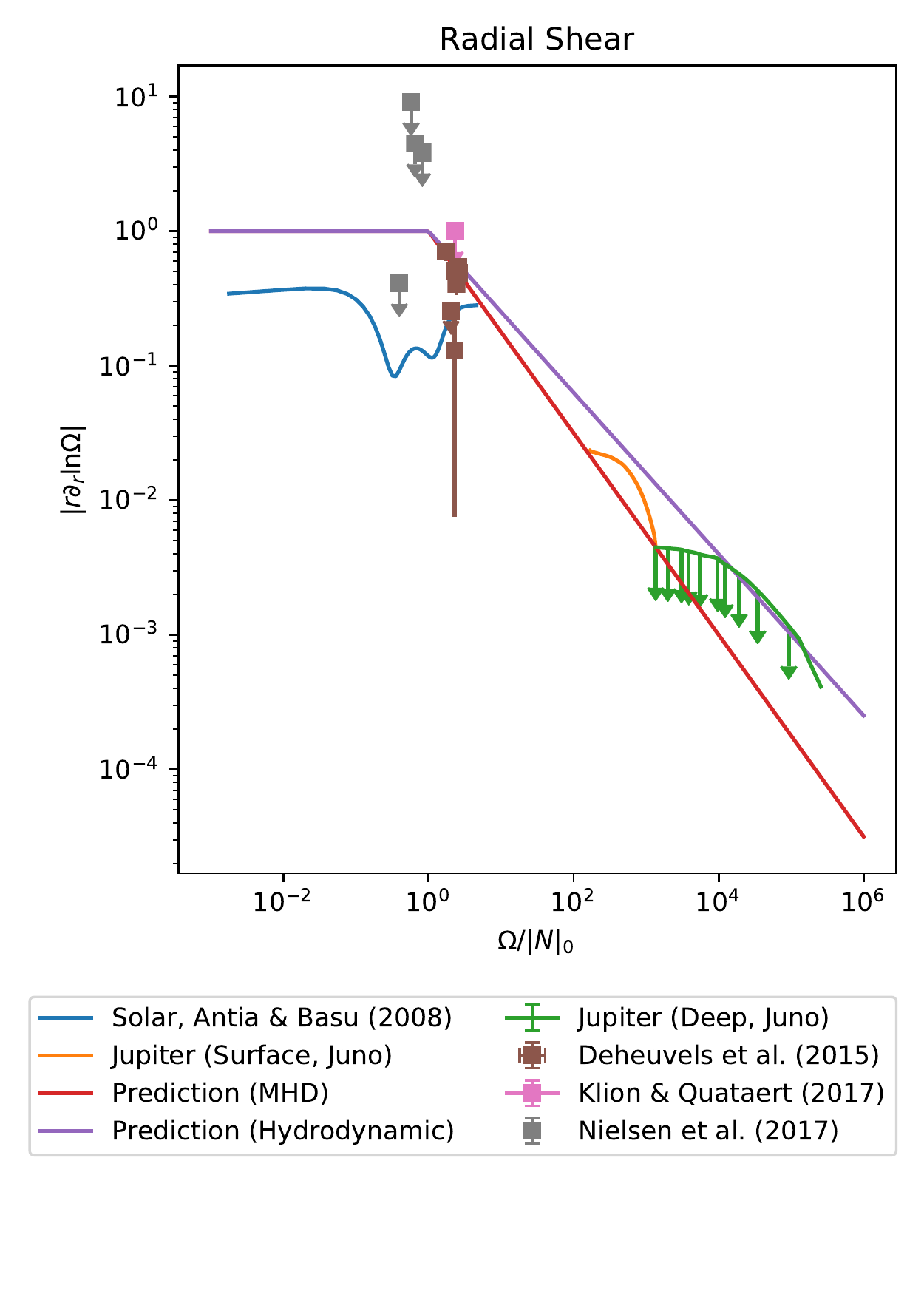}
\caption{The relative radial shear $|r\partial_r \ln\Omega|\approx |R\nabla\ln\Omega|$ is shown as a function of $\Omega/|N|_0$ for observed convecting stars and Jupiter alongside our prediction, which is constant for $\Omega < |N|_0$ and scales as $(\Omega/|N|_0)^{-3/4}$ (MHD) and $(\Omega/|N|_0)^{-3/5}$ (hydrodynamic) for $\Omega > |N|_0$. Squares denote asteroseismic results. Helioseismic data and constraints from {\emph{Juno}} are shown as solid lines. {\emph{Juno}} measurements of the shear at a depth of less than about $3000\rm{km}$ are shown separately from {\emph{Juno}} upper limits on that at greater depths.}
\label{fig:radial}
\end{figure}

We now turn to the (spherical) radial shear $|r \partial_r \ln \Omega|$.
In this case the observations come from a combination of helioseismology, asteroseismology and gravity multipole measurements of Jupiter (Fig.~\ref{fig:radial}).
The asteroseismology measurements are sensitive to the the total shear across the convection zone and so we have converted this into an average shear and plotted them at the average $|N|_0$ computed by equation~\eqref{eq:N_avg}.
The helioseismic and Juno results are plotted with profiles of $|N|_0$ from models of the Sun and Jupiter described in Appendices~\ref{appen:solar} and~\ref{appen:Jupiter}.

As before, the fastest rotator for which we have data is Jupiter~\citep{2018Natur.555..223K} and they show a preference in the outer layers for the hydrodynamic regime and in the inner layers for the MHD regime, though in both cases the preference is weak.
In the slowly rotating regime the solar data provide our main test and are broadly consistent with our prediction of no significant scaling~\citep{2001ApJ...559L..67A}.

The available asteroseismic measurements are mostly for slowly-rotating or moderately-rotating systems which lie close enough to the predicted break in the power-law ($\Omega \approx |N|$) that it is not clear which regime they probe.
Nonetheless they are broadly consistent with our expectations and show slight evidence of a trend of decreasing relative shear with increasing rotation rate.
Note that some caution is warranted in the interpretation of these data because the asteroseismic measurements of differential rotation probe a part of the star which often includes a stably stratified region.
There is thus some uncertainty in attributing the shear purely to the convection zone, though that is what we have done here.

The solar data as well as the bounds set by~\citet{2017A&A...603A...6N} and~\citet{2015A&A...580A..96D} suggest that the overall scale of our predictions is too large by a factor of $3$ or so.
This agrees with the offset we saw in the starspot measurements so we think it likely that the scale $\beta_{|R\nabla\Omega|} \approx 0.3$.%%%---------- close: radial_tests

%%%%%%%%% jump to xi_obs
%%%---------- open: xi_obs
\subsection{Baroclinicity}
\label{sec:xi_obs}

\begin{figure}
\centering
\includegraphics[width=0.47\textwidth]{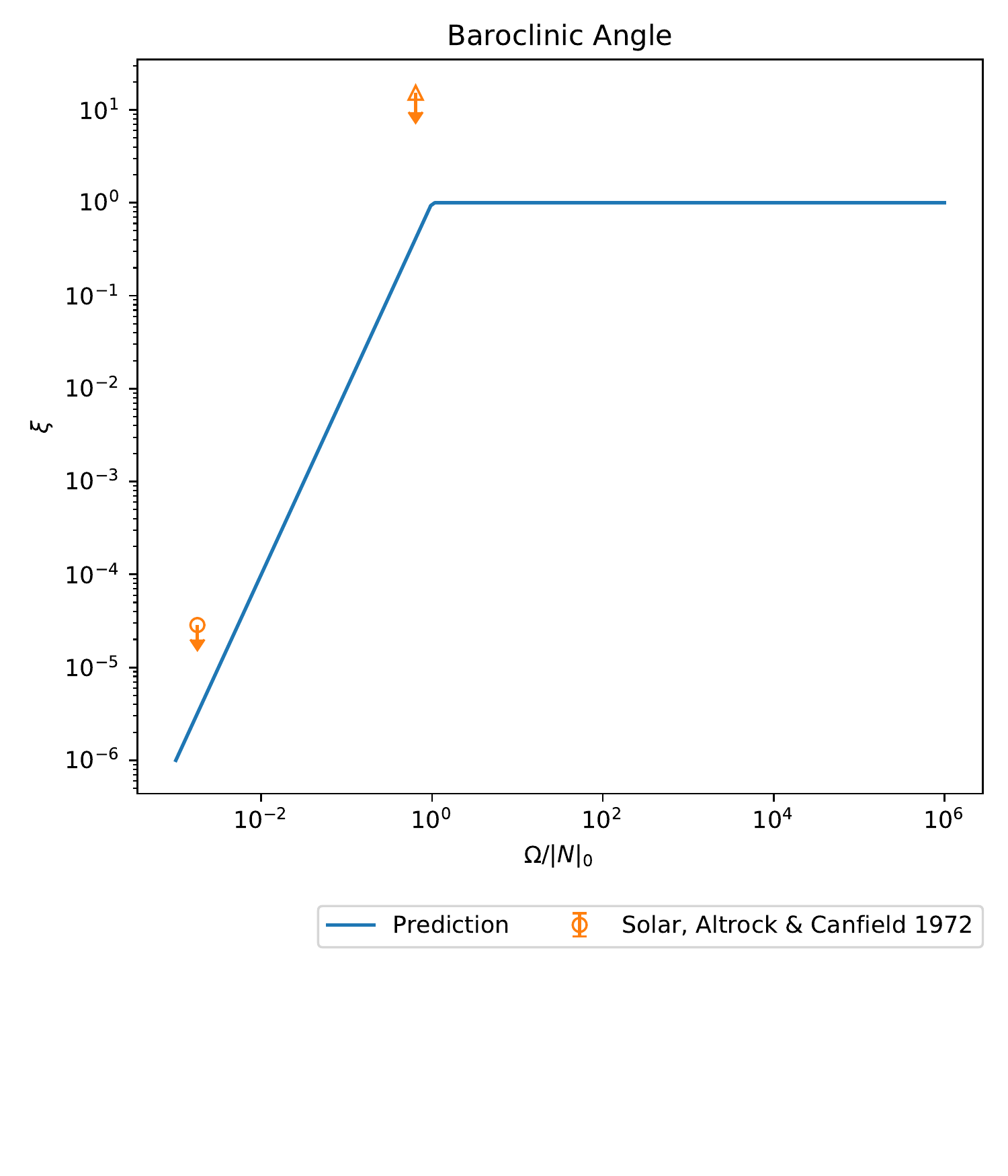}
\caption{The angle $\xi$ between the pressure gradient and the entropy gradient is shown from our predictions. It was also inferred for the Sun from the equator-pole temperature difference and the resulting upper bound is shown with a circle for the surface and a triangle for the inferred average~\citep{1972SoPh...23..257A}. Our prediction scales as $(\Omega/|N|_0)^2$ for slow rotation ($\Omega < |N|_0$) and is constant for rapid rotation ($\Omega > |N|_0$).}
\label{fig:xi_obs}
\end{figure}

In order to compute the differential rotation in convection zones we also needed to compute the angle $\xi$ between the pressure and entropy gradients.
This baroclinic angle is important because it determines the importance of the thermal wind term $\nabla p \times \nabla \rho$ in the vorticity equation.
It is also closely related to the temperature difference between the pole and the equator of a star because it is proportional to the temperature gradient along isobars.
Because of this, we may compare the observed temperature difference on the Sun to our predictions for $\xi$.

Unfortunately the pole-equator temperature difference on the Sun has proven difficult to measure due to significant observational uncertainties and the small expected signal~\citep{2017SoPh..292..123T}.
\citet{Rast_2008} summarizes a variety of efforts, most of which either conclude that the difference is zero to within uncertainties or provided a figure of order $1.5\,\mathrm{K}$.

Bearing in mind these uncertainties, in Fig.~\ref{fig:xi_obs} we have inferred $\xi$ from the measurements made by~\citet{1972SoPh...23..257A}, who found a temperature difference of $1.5\pm 0.6\,\mathrm{K}$.
We have translated done this with two different methods described in Appendix~\ref{appen:solar}.
In the first approach we assumed that $\xi$ is determined by the local properties of the convection zone, so we have plotted the resulting data at the local $\Omega/|N|_0$ near the surface of the solar convection zone.
In the second approach we assumed that the relative change in temperature between the pole and the equator persists throughout the convection zone and computed a volume-weighted average $\xi$ and an average $\Omega/|N|_0$, with the average \brvs\ frequency given by equation~\eqref{eq:N_avg}.

Bearing in mind the large uncertainties on this measurement, we find good agreement with our predicted $\Omega^2$ scaling for slow rotators.
The solar data do not probe the rapidly rotating limit so for that we rely on simulations in Section~\ref{sec:xi_sims}.
The scaling we find suggests that the baroclinic angle in convection zones is indeed driven by rotation-induced anisotropy in the convective heat flux~\citep{2018MNRAS.480.5427J}.%%%---------- close: xi_obs

%%%%%%%%% jump to activity
%%%---------- open: activity
\subsection{Magnetic Activity}
\label{activity}

Our prediction is that magnetic field strength is independent of rotation rate for $\Omega < |N|$ and increases with increasing rotation rate for $\Omega > |N|$.
This is somewhat at odds with observations of magnetic activity, which show increasing X-ray luminosity with increasing rotation rate for slowly rotating stars ($\Omega < |N|$) and a plateau for rapidly rotating stars~\citep[$\Omega > |N|$,][]{2011ApJ...743...48W}.
This poses two challenges.
First, if our prediction is correct for slowly-rotating stars, how can tracers of magnetic activity such as X-ray luminosity increase without the magnetic field strength increasing?
Secondly, if our prediction is correct for rapidly-rotating stars, how can tracers of activity plateau while the magnetic field strength increases?

We think the first challenge is most likely resolved by changes in the geometry of the magnetic field as a function of rotation rate.
In numerical simulations of convective dynamos the magnetic field becomes increasingly ordered on large scales as the rotation rate increases~\citep[see e.g.][]{doi:10.1111/j.1365-246X.2006.03009.x}.
If activity relies on the magnetic field having dipolar or low-order multipolar structure then the activity can increase with rotation rate even if the magnetic field strength is constant, as we predict for slowly-rotating stars.
This is supported by observations which suggest that the X-ray luminosity increases with increasing large-scale magnetic field~\citep[][figure~6]{10.1093/mnras/stu728}.

The second challenge could be explained in a few different ways.
First, surface convection zones are often inefficient, meaning that most of the heat flux is carried radiatively rather than by advection.
As a result increasing the rotation rate would not increase $|N|$ as we have predicted in the case of efficient convection zones.
Because we predict that $\varv_{\rm A} \approx h |N|$ this suggests that, even when $\Omega \gg |N|$, the surface magnetic field strength does not increase with increasing rotation rate.
The entire effect of increasing activity with increasing rotation rate would then be due to the field geometry becoming dipolar.
This eventually saturates around $\Omega \approx |N|$, producing a plateau.
Note that this explanation requires that the surface magnetic field be generated near the surface, such that it is sensitive to the properties of near-surface convection.

A further possibility is that the magnetic activity depends primarily on a spot coverage fraction which saturates and so no further increase in field strength can produce more activity~\citep{1984A&A...133..117V}.
Testing such a proposition is difficult, but if it holds it would resolve the apparent conflict between our predictions and the observations.%%%---------- close: activity
%%%---------- close: activity

%%%%%%%%%%% jump to simulations
%%%---------- open: simulations
\section{Numerical Simulations}
\label{sec:sims}

We now turn to numerical simulations.
Because these typically report the Rayleigh number, we can directly compute $|N|$ and so we rephrase our scaling relations in terms of the \brvs\ frequency which is actually realized in the system, rather than the non-rotating one we used in section~\ref{sec:obs}.
The reason that $|N|$ is not the same as $|N|_0$ is that rotation generally makes convection less efficient at transporting heat.
This means that the entropy gradient has to be steeper for convection to carry the same heat flux in a rotating system than in a non-rotating one.

The relation between $|N|$ and $|N|_0$ is derived in detail in Paper~I.
Briefly, in non-magnetized convection zones, we compute the heat flux as
\begin{align}
F \approx \frac{\rho c_{\rm p} T}{\mu} D |\nabla s|,
\end{align}
where $D \approx \varv_{\rm c}^2/|N|$ is the turbulent diffusivity.
Inserting our predicted scaling for $\varv_{\rm c} \propto |N|^{3/2} \Omega^{-1/2}$ then yields
\begin{align}
F \approx \rho h^3 |N|^5 \Omega^{-2}.
\end{align}
Setting $F$ to be constant then yields $|N| \propto \Omega^{2/5}$.
The analysis is similar in magnetized convection zones, but begins with
\begin{align}
F \approx \rho \varv_{\rm c} \varv_{\rm A}^2.
\end{align}
Inserting our scaling for $\varv_{\rm c}$ and the scalings $\varv_{\rm A} \approx h |N|$ yields
\begin{align}
F \approx \rho h^3 |N|^4 \Omega^{-1}
\end{align}
and so $|N| \propto \Omega^{1/4}$, as given in Table~\ref{tab:summary}.

\begin{table*}
\caption{The scalings of the differential rotation, meridional circulation, baroclinicity, convective velocity, and the ratio of magnetic to kinetic energy are given for the three regimes of interest. Note that the latitudinal and spherical radial differential rotation are each formed of a mixture of the cylindrical vertical and radial differential rotation. Because the cylindrical radial shear is larger than the vertical shear, both spherical components of the differential rotation share the scaling of the former.}
\label{tab:summary_num}
\begin{tabular}{llllll}
\hline
\hline
Case & $\frac{|R\nabla \Omega|}{\Omega}$ & $\frac{|R\partial_R \Omega|}{\Omega}$ & $\frac{|R\partial_z \Omega|}{\Omega}$ & $\frac{|r\partial_r \Omega|}{\Omega}$ &$\frac{|\partial_\theta \Omega|}{\Omega}$\\
\hline
Slow ($\Omega \ll |N|_0$) & $1$ & $1$ & $1$ & $1$ & $1$\\
Fast Hydro.($\Omega \gg |N|$) & $\left(\frac{\Omega}{|N|}\right)^{-1}$& $\left(\frac{\Omega}{|N|}\right)^{-1}$ & $\left(\frac{\Omega}{|N|}\right)^{-2}$& $\left(\frac{\Omega}{|N|}\right)^{-1}$& $\left(\frac{\Omega}{|N|}\right)^{-1}$\\
Fast MHD ($\Omega \gg |N|$) &$\left(\frac{\Omega}{|N|}\right)^{-1}$&$\left(\frac{\Omega}{|N|}\right)^{-1}$& $\left(\frac{\Omega}{|N|}\right)^{-2}$ &$\left(\frac{\Omega}{|N|}\right)^{-1}$&$\left(\frac{\Omega}{|N|}\right)^{-1}$\\
\hline
\hline
Case & $\frac{u_r}{h|N|}$ & $\frac{u_\theta}{h|N|}$ & $\xi$ &$\frac{\varv_{\rm c}}{h|N|}$ & $\frac{\varv_{\rm A}^2}{\varv_{\rm c}^2}$\\
\hline
Slow ($\Omega \ll |N|$) & $\frac{h}{r}\left(\frac{\Omega}{|N|}\right)^{2}$ & $\left(\frac{\Omega}{|N|}\right)^{2}$ & $\left(\frac{\Omega}{|N|}\right)^2$ & $1$ & $1$\\
Fast Hydro.($\Omega \gg |N|$) & $\frac{h}{r}\left(\frac{\Omega}{|N|}\right)^{-3}$& $\left(\frac{\Omega}{|N|}\right)^{-3}$ & $1$  & $\frac{|N|}{\Omega}$ & N/A\\
Fast MHD ($\Omega \gg |N|$) & $\frac{h}{r}\left(\frac{\Omega}{|N|}\right)^{-1}$& $\left(\frac{\Omega}{|N|}\right)^{-1} \approx\left(\frac{\Omega}{|N|}\right)^{-1}$ & $1$  & $\frac{|N|}{\Omega}$ & $\frac{\Omega^2}{|N|^2}$\\
\hline
\hline
\end{tabular}
\end{table*}

Our predictions in terms of $|N|$ are given in Table~\ref{tab:summary_num}.
Once more we emphasize that we have only predicted the scalings of these various quantities but not their actual magnitudes.
We expect that in each case there are likely factors of order unity in these relations, such that for a quantity $Q$ scaling as $(\Omega/|N|)^\alpha$ we have
\begin{align}
	Q = \lambda_Q Q_0 \left(\frac{\Omega}{|N|}\right)^\beta
\end{align}
for some $\lambda_Q$ of order unity and dimensional $Q_0$ which gives the characteristic scale in terms of $h$ and $|N|$.
For clarity we have non-dimensionalized all quantities appearing in Table~\ref{tab:summary}, so that in each case $Q_0 = 1$.

We now test the relations in Table~\ref{tab:summary_num}.
Details of how the data from various sources were obtained, standardized and processed are provided in Appendix~\ref{appen:data}, and all scripts and files needed to reproduce this analysis are available on \url{Zenodo.org}.

%%%%%%%%% jump to hydro_sim_tests
%%%---------- open: hydro_sim_tests

\subsection{Hydrodynamic Simulations}
\label{sec:hydro}

%%% Mention predicted slopes in captions

\begin{figure}
\centering
\includegraphics[width=0.47\textwidth]{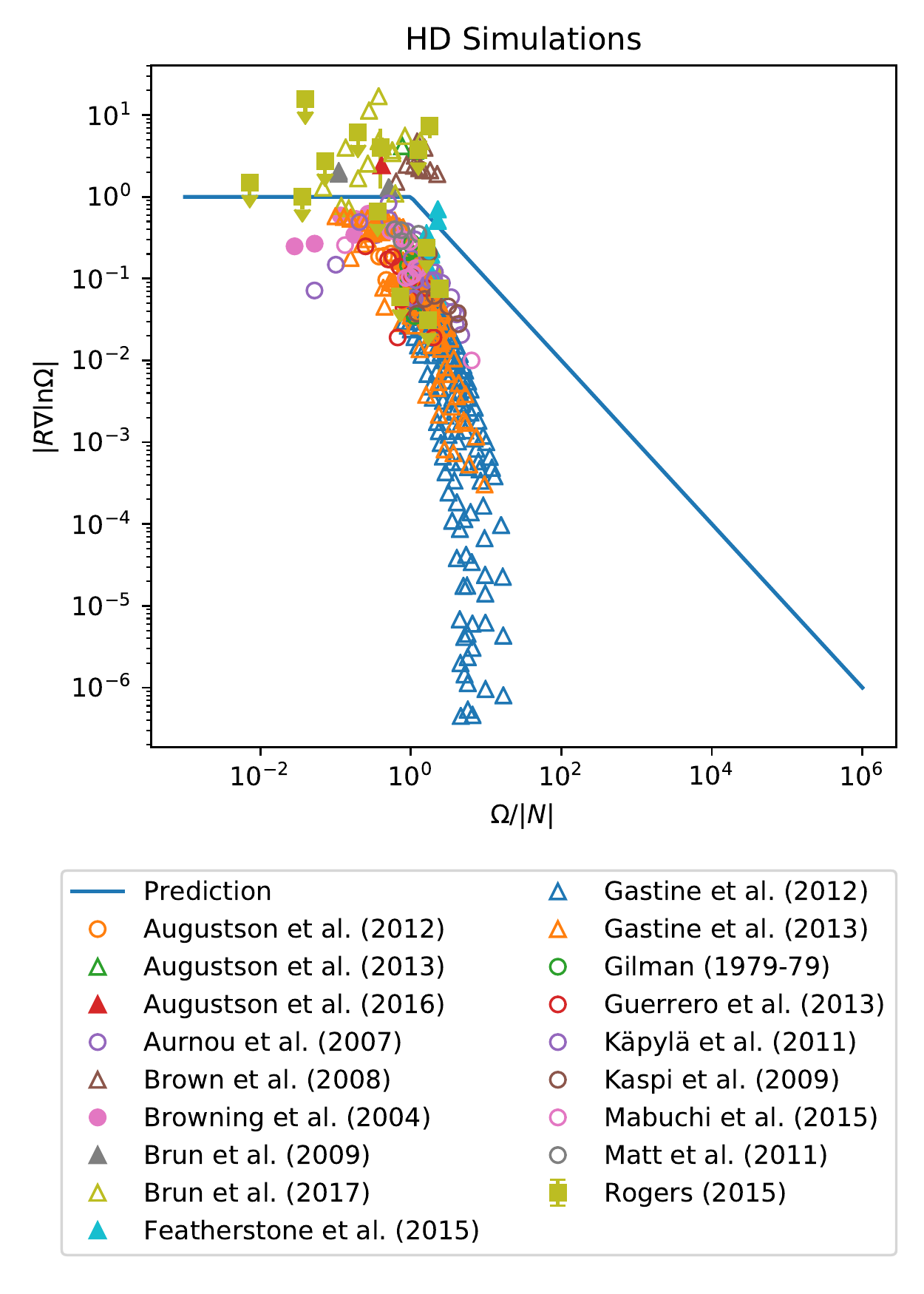}
\caption{The relative differential rotation $|R\nabla\ln\Omega|$ is shown alongside our prediction as a function of $\Omega/|N|$ for a variety of hydrodynamic convection simulations. Squares indicate radial shear, circles indicate latitudinal shear and triangles indicate the root-mean square of the shear integrated over the domain. Filled shapes indicate simulations performed on deep spherical shell domains while open shapes indicate those performed in shellular domains. Our prediction is constant for slow rotation ($\Omega < |N|$) and scales as $(\Omega/|N|)^{-1}$ for rapid rotation ($\Omega > |N|$).}
\label{fig:hydro}
\end{figure}

Because simulations tend to probe different regions of parameter space they provide a complementary set of tests.
In this case simulations tend to fill in the slowly-rotating and moderately-rotating limits more than the observations and this allows us to test whether the scaling indeed changes between these limits.

The results of a variety of different hydrodynamic convection simulations performed for different geometries, with different software instruments, in both the anelastic\footnote{The anelastic approximation is the limit as the Mach number and Froude number both go to zero, meaning that characteristic velocities are much less than both the sound speed and the free-fall speed across the domain~\citep{a6ec03cc63c04d33a070ecaa1bb621d5}. This allows the pressure and gravity terms in the Navier-Stokes equation to be written in terms of an entropy perturbation away from a background state, and imposes the constraint that $\nabla\cdot(\bar{\rho}\boldsymbol{v})=0$, where $\bar{\rho}$ is the background density profile.} and fully compressible limits, and with different choices of dimensionless parameters are shown in Fig.~\ref{fig:hydro}.
Squares indicate radial shear, circles indicate latitudinal shear and triangles indicate the root-mean square of the shear integrated over the domain. Filled shapes indicate simulations performed on deep spherical shell domains\footnote{These are domains which just exclude a small region in the centre of the sphere for numerical or algorithmic reasons.}, which are meant to mimic full spheres, while open shapes indicate those performed in thinner shellular domains.

\begin{figure}
\centering
\includegraphics[width=0.47\textwidth]{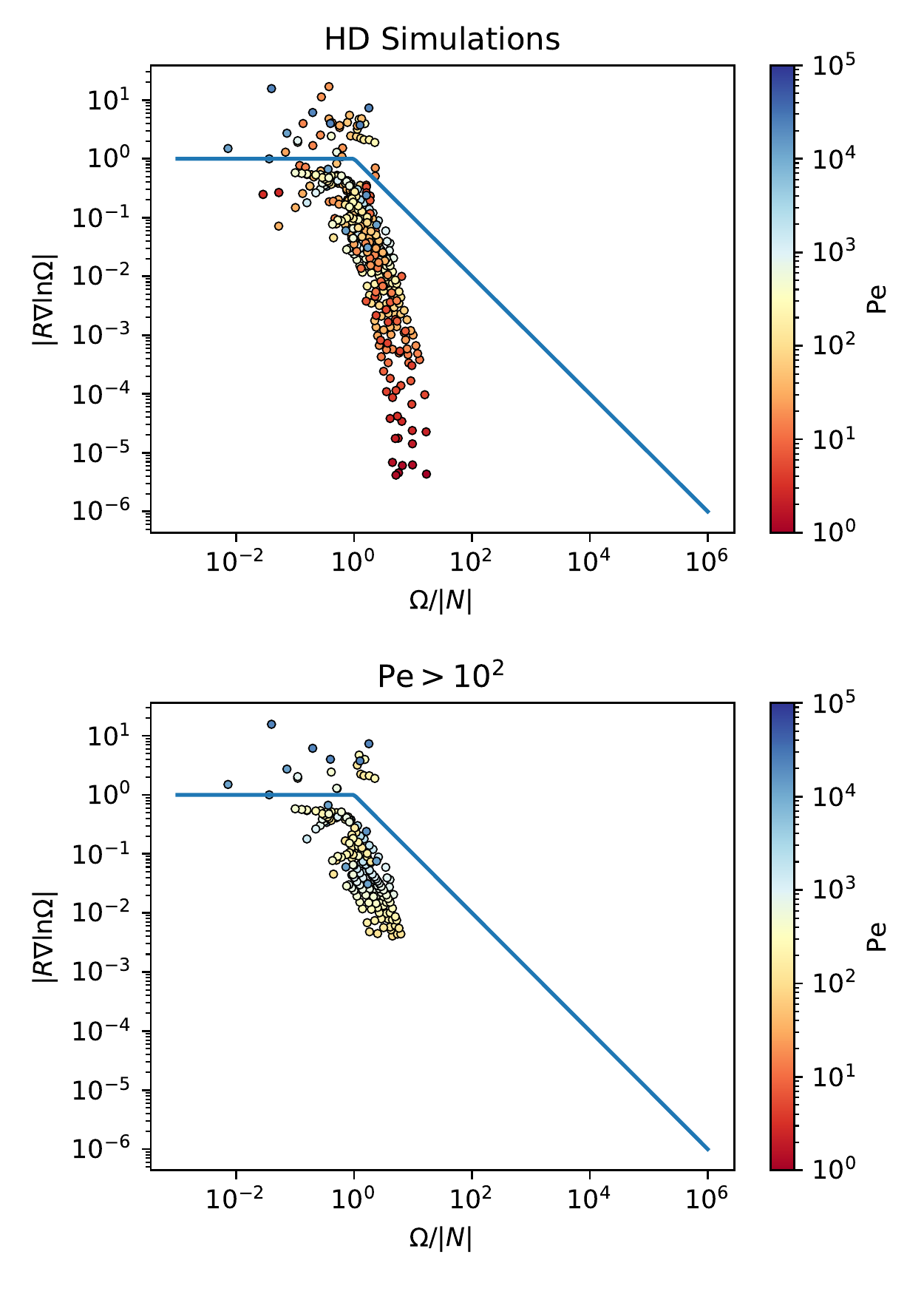}
\caption{The relative differential rotation $|R\nabla\ln\Omega|$ is shown alongside our prediction as a function of $\Omega/|N|$ for a variety of hydrodynamic convection simulations, coloured by P\'{e}clet number $\mathrm{Pe}$. The upper panel shows data from all simulations for which we could compute $\mathrm{Pe}$ while the lower shows just those simulations with $\mathrm{Pe} > 10^2$. The slopes of the power-law predictions are indicated in the relevant regimes.}
\label{fig:hydro_pe}
\end{figure}

There is significant scatter between simulations performed by different groups.
Some of this may result from the use of different software instruments, though several of the instruments used do yield identical results on identical problems~\citep{2011Icar..216..120J}
It could also be due to imperfections in our standardization of the data.
In particular, different symbols in Fig.~\ref{fig:hydro} correspond to different ways in which the differential rotation was reported, and so reflect different kinds of spatial averaging as well as different weightings of the radial and latitudinal components of $\nabla\Omega$.
We do not see any systematic trends across different reported measures, though, so this also seems unlikely to be the cause of a majority of the scatter.

There is also some scatter within the data from individual sources.
Different choices of boundary conditions or geometry can contribute to this scatter~\citep[see e.g.][]{2016GeoJI.204.1120Y}, but it seems likely that a large fraction of it results from the use of different Prandtl, Rayleigh or Reynolds numbers, and we expect such differences to become smaller as the latter two numbers become larger.

To probe these effects across simulations we use the P\'{e}clet number
\begin{align}
	\mathrm{Pe} \equiv \frac{\varv d}{\alpha},
\end{align}
where $\alpha$ is the thermal diffusivity, $d$ is a characteristic length scale, $\varv$ is the root-mean square of the velocity in the rotating frame with zero net angular momentum.
When the P\'{e}clet number is sufficiently large convection is efficient because the heat transport comes to be dominated by advection rather than diffusion.
We choose this number despite it being an \emph{output} from simulations rather than an input because it directly measures whether or not the convection is efficient and this is a critical factor in determining whether or not our predictions are applicable.

The upper panel of Fig.~\ref{fig:hydro_pe} shows the same simulations coloured by P\'{e}clet number\footnote{For details of how this was calculated for each simulation see Appendix~\ref{appen:data}.}.
There is a clear trend towards less shear with lower $\mathrm{Pe}$.
In the lower panel we retain only those simulations with $\mathrm{Pe} > 10^2$.
Doing so significantly reduces the scatter and improves the agreement with our predictions.

Some of the remaining scatter is due to density stratification.
In particular, the different tracks that can be seen in the data of~\citet{2012Icar..219..428G} in Fig.~\ref{fig:hydro} correspond to different degrees of density stratification.
The strength of this effect decreases with increasing P\'{e}clet number.
This could be because, as the density stratification increases, the velocity required to carry the heat flux becomes large near the top of the domain and small near the bottom.
So when the mean $\mathrm{Pe}$ remains constant, increasing stratification means that an increasingly large proportion of the domain has a low $\mathrm{Pe}$.
As such more work may be required to probe whether this effect holds at the large $\mathrm{Pe} > 10^6$ that apply in the Sun for instance.

If there is an effect of density stratification at large $\mathrm{Pe}$, it could be due to the compressibility torque described by~\citet{2009GApFD.103...31G}.
Such effects are potentially very physically important if they hold in the astrophysically relevant high-$\mathrm{Pe}$ limit particularly because some systems, such as red giants, exhibit extreme degrees of density stratification in their convection zones.

Focusing just on the high-$\mathrm{Pe}$ number data shown in the lower panel of Fig.~\ref{fig:hydro_pe}, we see an overall trend for the relative shear to plateau towards the slow rotation limit and to fall in the rapidly rotating limit.
The existence of the plateau suggests that the scaling of the off-diagonal components of the Reynolds stress is indeed as we have argued in Part~I, with the $r\theta$ component scaling quadratically and the $r\phi$ and $\theta\phi$ components scaling linearly.

In the rapidly rotating limit our scaling appears to approximately bound the results of simulations.
That is, some simulations report shears significantly below our predictions while very few report anything significantly above them and the largest shears found in simulations appear to track our scaling reasonably well.

\begin{figure}
\centering
\includegraphics[width=0.47\textwidth]{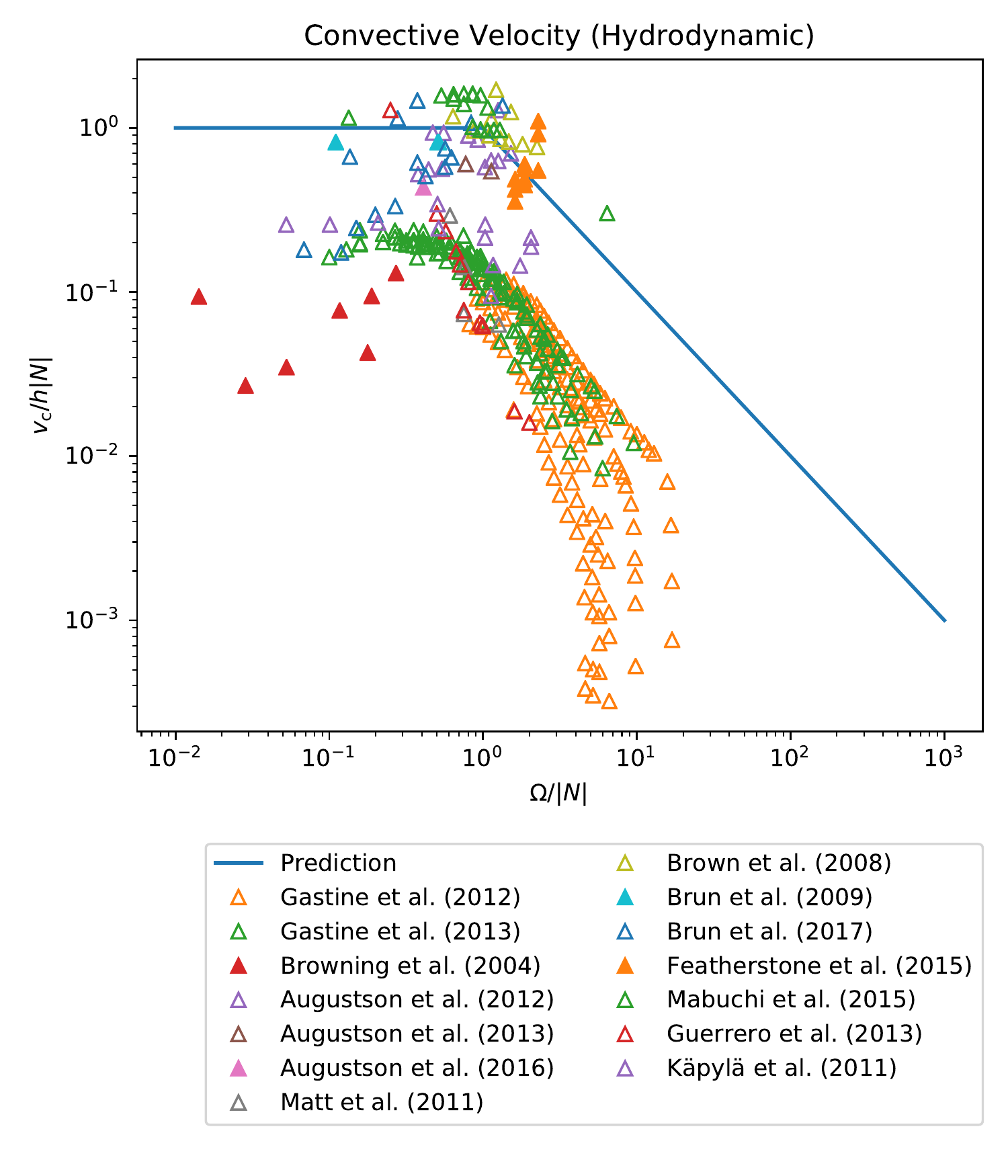}
\caption{The normalized convection speed $\varv_{\rm c} / h |N|$ is shown alongside our predictions as a function of $\Omega/|N|$ for a variety of hydrodynamic convection simulations. Filled shapes indicate simulations performed on spherical domains while open shapes indicate those performed in shellular domaints. Triangles denote outputs which were averaged over the simulation domain. Our prediction is constant for slow rotation ($\Omega < |N|$) and scales as $(\Omega/|N|)^{-1}$ for rapid rotation ($\Omega > |N|$).}
\label{fig:vc_HD}
\end{figure}

A second test of our predictions comes from the convective velocities.
Fig.~\ref{fig:vc_HD} shows our predictions for the ratio $\varv_{\rm c} / h |N|$ alongside the results of the subset of these simulations which reported enough information for us to calculate this ratio.
Note that we have used $|N|$ rather than $|N|_0$ in the denominator to emphasize the suppression relative to the predictions of non-rotating mixing length theory.

\begin{figure}
\centering
\includegraphics[width=0.47\textwidth]{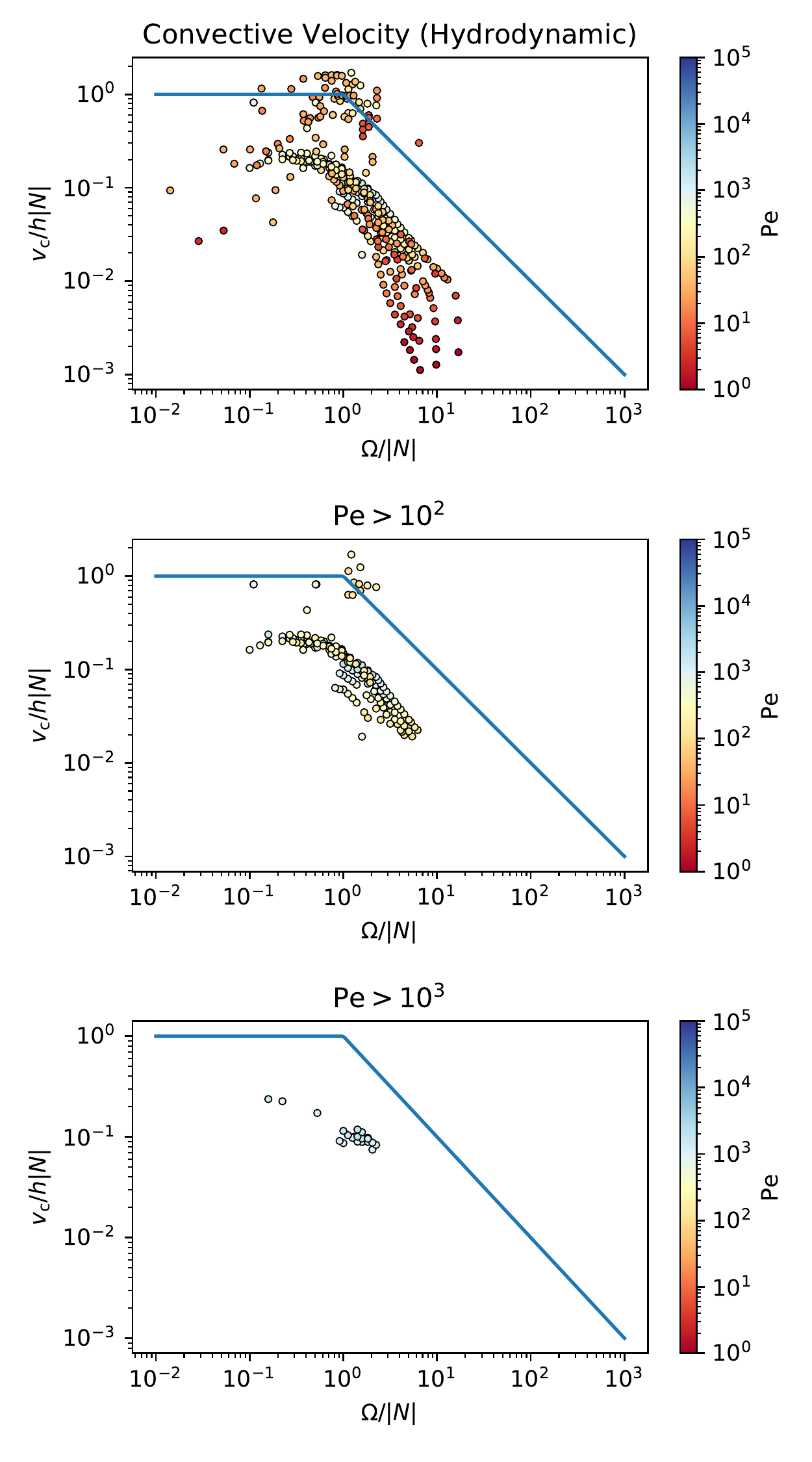}
\caption{The normalized convection speed $\varv_{\rm c} / h |N|$ is shown alongside our predictions as a function of $\Omega/|N|$ for a variety of hydrodynamic convection simulations, coloured by P\'{e}clet number $\mathrm{Pe}$. The upper panel shows data from all simulations for which we could compute $\mathrm{Pe}$, the lower shows just those simulations with $\mathrm{Pe} > 10^2$, and the lower panel shows just those with $\mathrm{Pe} > 10^3$.  Our prediction is constant for slow rotation ($\Omega < |N|$) and scales as $(\Omega/|N|)^{-1}$ for rapid rotation ($\Omega > |N|$).}
\label{fig:vc_HD_Pe}
\end{figure}

We see that the data all exhibit the same trend of a plateau towards slow rotation and a decline towards rapid rotation.
There is again considerable scatter between simulations performed by different groups.
This is again primarily due to many of these simulations having a low P\'{e}clet number.
We can see this in Fig.~\ref{fig:vc_HD_Pe}, which shows the full suite of simulations as well as cuts with $\mathrm{Pe} > 10^2$ and $\mathrm{Pe} > 10^3$.
As we restrict the P\'{e}clet number to be ever-larger the scatter reduces and the agreement with our predictions improves.

Note that we tend to overestimate the convection speed relative to the simulations.
This is just because we have set all dimensionless constants to unity in our theory.
In practice we can absorb the apparent constant offset into our predictions by setting the prefactor $\lambda_{\varv_{\rm c}} \approx 0.2$ in our relations.

Finally, while we have focused on simulations in spherical geometries, our results also agree with the cartesian simulations of~\citet{2014ApJ...791...13B}, who find, as we do, that $\varv_{\rm c} \propto \Omega^{-1/5}$ at fixed heat flux.

%%%---------- close: hydro_sim_tests

%%%%%%%%% jump to mhd_sim_tests
%%%---------- open: mhd_sim_tests

\subsection{MHD Simulations}
\label{sec:mhd}

%%% Mention predicted slopes in captions

\begin{figure}
\centering
\includegraphics[width=0.47\textwidth]{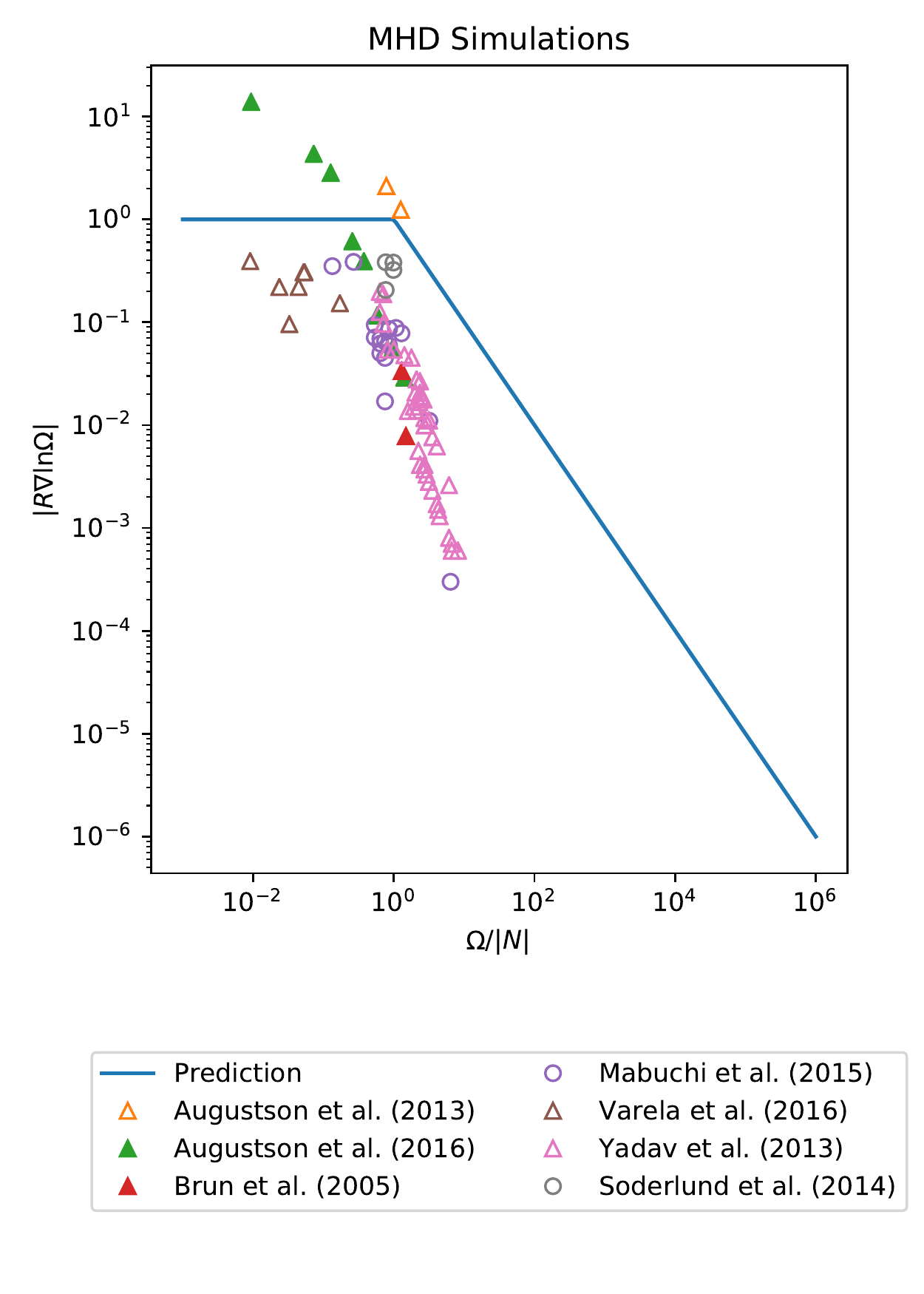}
\caption{The relative differential rotation $|R\nabla\ln\Omega|$ is shown alongside our predictions as a function of $\Omega/|N|$ for a variety of MHD convection simulations. Circles indicate latitudinal shear and triangles indicate root-mean-squared shear integrated over the domain. Filled shapes indicate simulations performed on spherical domains while open shapes indicate those performed in shellular domaints. Our prediction is constant for slow rotation ($\Omega < |N|$) and scales as $(\Omega/|N|)^{-1}$ for rapid rotation ($\Omega > |N|$).}
\label{fig:mhd}
\end{figure}

\begin{figure}
\centering
\includegraphics[width=0.47\textwidth]{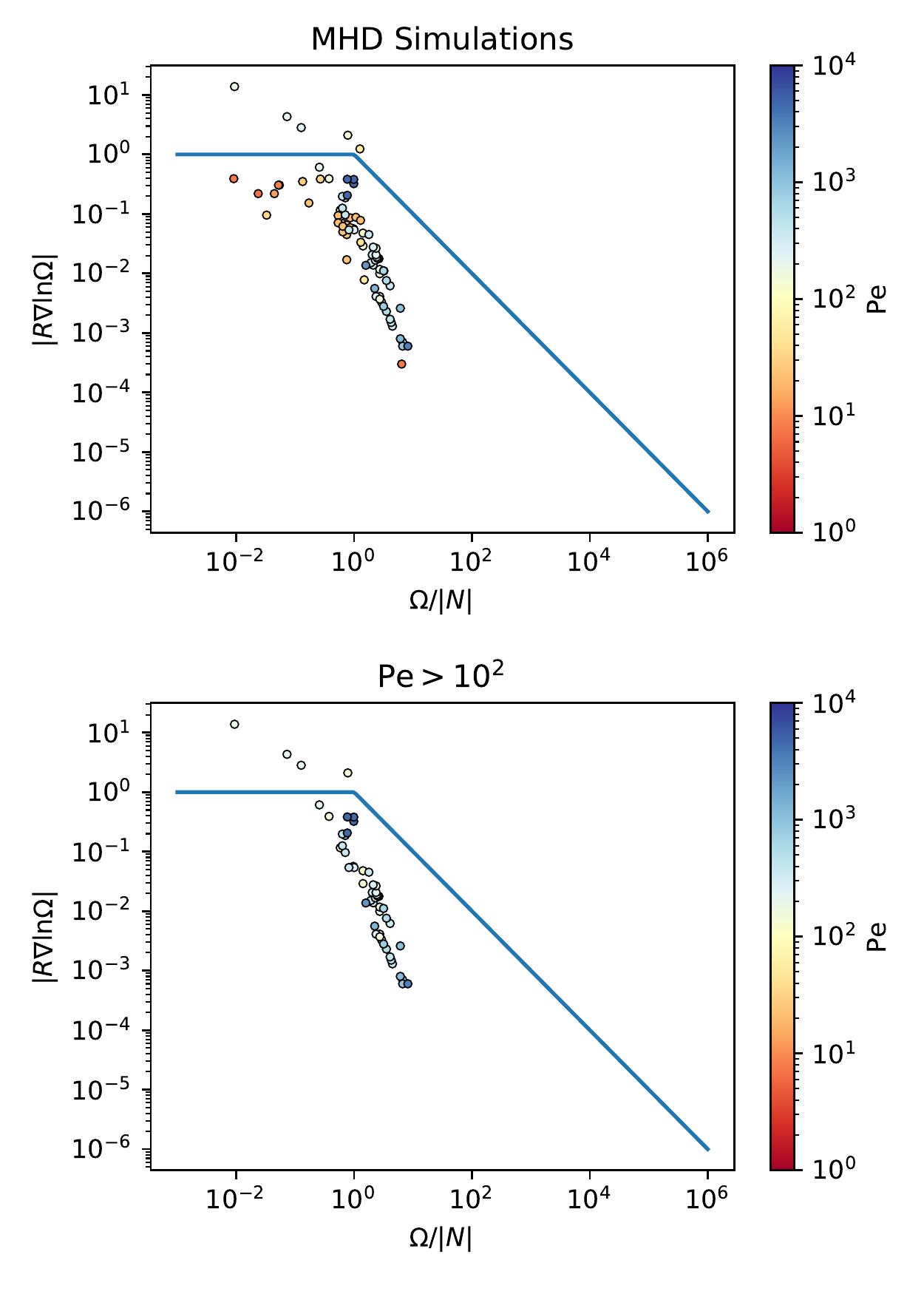}
\caption{The relative differential rotation $|R\nabla\ln\Omega|$ is shown alongside our predictions as a function of $\Omega/|N|$ for a variety of MHD convection simulations, coloured by P\'{e}clet number $\mathrm{Pe}$. The upper panel shows data from all simulations for which we could compute $\mathrm{Pe}$ while the lower shows just those simulations with $\mathrm{Pe} > 10^2$. Our prediction is constant for slow rotation ($\Omega < |N|$) and scales as $(\Omega/|N|)^{-1}$ for rapid rotation ($\Omega > |N|$).}
\label{fig:mhd_pe}
\end{figure}

While hydrodynamic systems provide a helpful test of our arguments, the most common astrophysical case is the near-ideal MHD limit.
Fig.~\ref{fig:mhd} shows our predicted differential rotation scaling along with the findings of various MHD simulations of convection.
As before we include both anelastic and fully compressible simulations.
We employ the same conventions that filled shapes denote spherical domains, open shapes denote shellular domains, squares indicate radial shear, circles indicate latitudinal shear and triangles indicate root-mean square of the shear integrated over the domain.

There is much less scatter between simulations performed by different groups than in the case of hydrodynamic simulations.
This is largely because most of our data come from simulations performed with the \code{ASH} software instrument and that these data have been reported in a relatively uniform way.

There is also less scatter between the simulations in each individual work.
We believe this is partly because these simulations were performed using a smaller set of software instruments and partly because they used a more uniform set of dimensionless parameters\footnote{The greater uniformity of dimensionless parameters in these simulations is likely because the simulation must be fairly turbulent (advection-dominated) to see a dynamo in the first place, which effectively sets a floor on the P\'e{clet} number.}.
We can see the latter in a plot coloured by P\'e{clet} number, shown in the upper panel of Fig.~\ref{fig:mhd_pe}.
There are far fewer simulations at very low $\mathrm{Pe}$ than in the hydrodynamic sample and this results in less scatter.
When we restrict the sample to just $\mathrm{Pe} > 10^2$ (lower panel) the scatter goes away almost entirely and the remaining simulations lie nicely on our predicted slope, albeit out to slower rotation rates than we expect.

Overall we see a steeper slope than our predicted scaling in the rapidly rotating limit.
This conclusion is driven mostly by the results of~\citet{2013Icar..225..185Y} and~\citet{2016ApJ...829...92A}, which suggest that $|R\nabla\ln\Omega| \propto \Omega^{-2}$ rather than our predicted $\Omega^{-1}$.
This is consistent with the power-law fits provided by~\citet{2013Icar..225..185Y} for the zonal (azimuthal) and non-zonal (meridional) Rossby number\footnote{See also~\citet{10.1093/gji/ggt167} for more discussion of these scaling relations.}.
We may translate into predictions for the differential rotation at fixed heat flux\footnote{$|R\nabla\ln \Omega|\approx \mathrm{Ro}_{\rm zonal}$. Using $\mathrm{Ra}_Q^* \propto \Omega^{-3}$ we then obtain the scaling for $|R\nabla\ln\Omega|$.}, suggesting that $|R\nabla\ln\Omega| \propto \Omega^{-1.32}$\footnote{Using their fit for dipolar dynamos with no $\mathrm{Pm}$ dependence.}, $\Omega^{-1.44}$\footnote{Using their fit for dipolar dynamos with a $\mathrm{Pm}$ dependence.}, $\Omega^{-1.2}$\footnote{Using their fit for multipolar dynamos with no $\mathrm{Pm}$ dependence.} or $\Omega^{-1.29}$\footnote{Using their fit for multipolar dynamos with a $\mathrm{Pm}$ dependence.}.~\citet{2013ApJ...774....6Y} similarly find a scaling of $\Omega^{-1.35}$.

Such steep power laws appears inconsistent with the observations shown in Figs.~\ref{fig:latitudinal} and~\ref{fig:radial} so we remain uncertain as to whether the slope in nature is truly steeper than our prediction.
However, if this is indeed what happens it points to a difficulty in our analysis.
In Section~7 of Paper~I we assumed that each of the shear, circulation and baroclinicity are as large as allowed by the conditions of heat and momentum balance.
While it must be that in each equation at least one of these saturates its bounds, it need not be the case that they all do.
If they do not then our predictions in Tables~\ref{tab:summary} and~\ref{tab:summary_num} are really upper bounds, only one of which must be saturated in any given scenario.
So for instance it could be that the cylindrical vertical differential rotation $\partial_z \Omega$ saturates our bound and scales like $\Omega^{-2}$, similar to what we see in Fig.~\ref{fig:mhd}.
In that case there is no formal requirement that the radial shear $\partial_R\Omega$ also saturates its bound, in which case the shear only needs to be as large as $\partial_z \Omega \propto \Omega^{-2}$.

In the slowly rotating regime the simulations of~\citet{2016AdSpR..58.1507V} hint at a plateau but those of~\citet{2016ApJ...829...92A} instead show a continuing increase of relative shear with decreasing rotation rate.
This could indicate that we have misplaced the break point between slow and rapid rotation.
On the other hand, all of the simulations which show this trend were performed on a spherical domain and so it could be that, even though the convective turnover is fast relative to rotation on average, deep in the core where convection is slow we might still be in the rapidly rotating limit.

\begin{figure}
\centering
\includegraphics[width=0.47\textwidth]{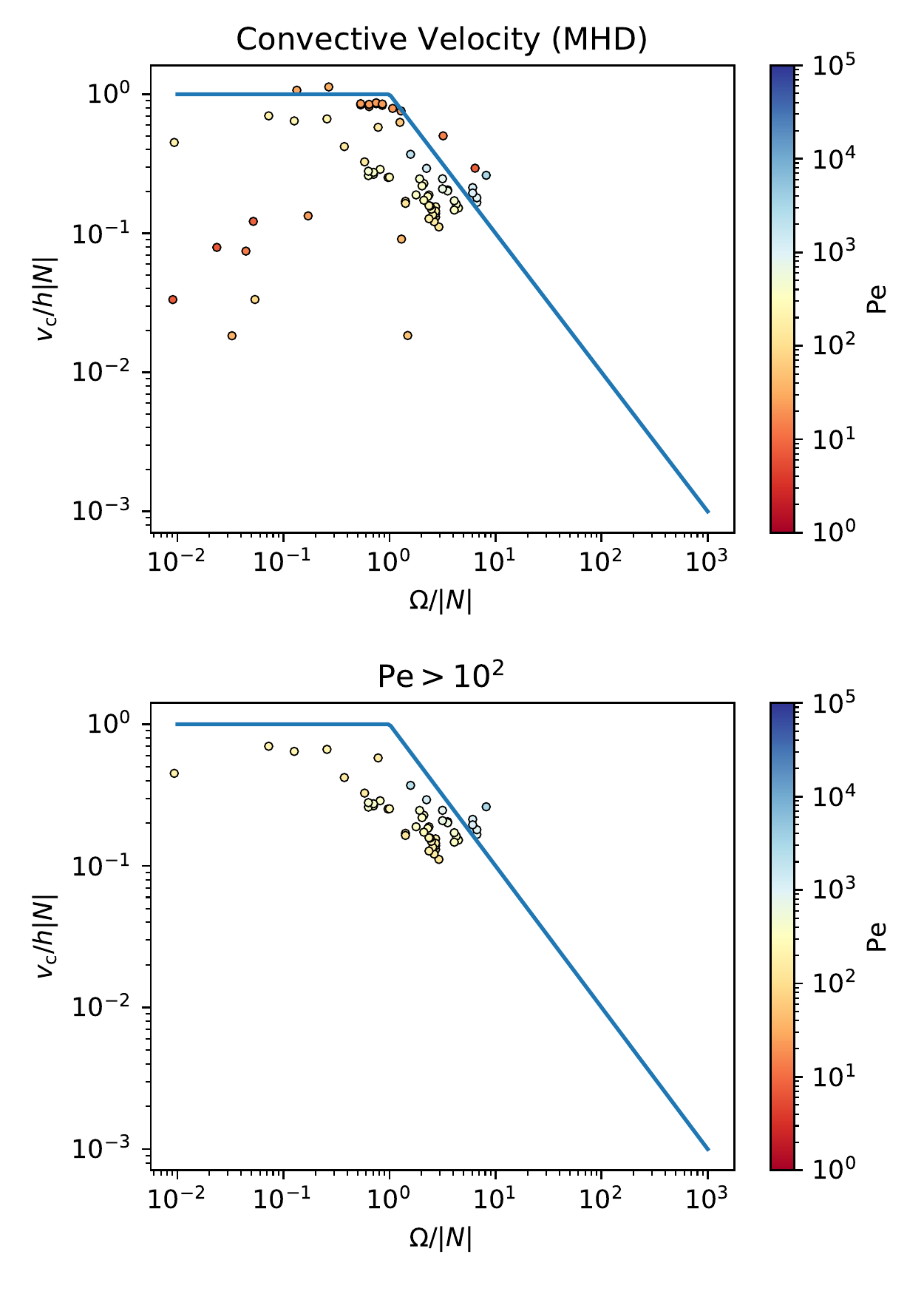}
\caption{The normalized convection speed $\varv_{\rm c} / h |N|$ is shown alongside our predictions as a function of $\Omega/|N|$ for a variety of MHD convection simulations, coloured by P\'{e}clet number $\mathrm{Pe}$. Our prediction is constant for slow rotation ($\Omega < |N|$) and scales as $(\Omega/|N|)^{-1}$ for rapid rotation ($\Omega > |N|$).}
\label{fig:vc_MHD}
\end{figure}

A further test of our predictions comes from the convective velocities.
Fig.~\ref{fig:vc_MHD} shows our predictions for the ratio $\varv_{\rm c} / h |N|$ alongside the results of the subset of these simulations for which we were able to calculate this ratio.

We see that the data generally exhibit the same trend of a plateau towards slow rotation and a decline towards rapid rotation.
Most of the scatter in these results is directly attributable to not being in the limit of efficient convection.
We can see this in the lower panel where we have filtered for $\mathrm{Pe} > 10^2$ and see significantly better agreement with our predicted slope and plateau.

Other authors have found similar, though typically weaker, scalings of MHD convection speed at constant heat flux.
We predict $\varv_{\rm c} \propto \Omega^{-1/2}$, while others find $\Omega^{-0.23}$ to $\Omega^{-0.29}$~\citep{doi:10.1111/j.1365-246X.2006.03009.x}\footnote{See their equations (30) and (31).}, $\Omega^{-0.41}$~\citep{2013Icar..225..185Y}\footnote{From their multipolar fit with magnetic Prandtl number dependence.} and $\Omega^{-0.32}$ to $\Omega^{-0.47}$~\citep{2017JFM...813..558A}\footnote{See their figure 11(a). Their $\epsilon$ is proportional to $\Omega^{-3}$.}.
This weaker scaling could be a result of the P\'e{clet} number dependence we have noted, though of course it could also be that our theory predicts a steeper scaling law than actually occurs.

\begin{figure}
\centering
\includegraphics[width=0.47\textwidth]{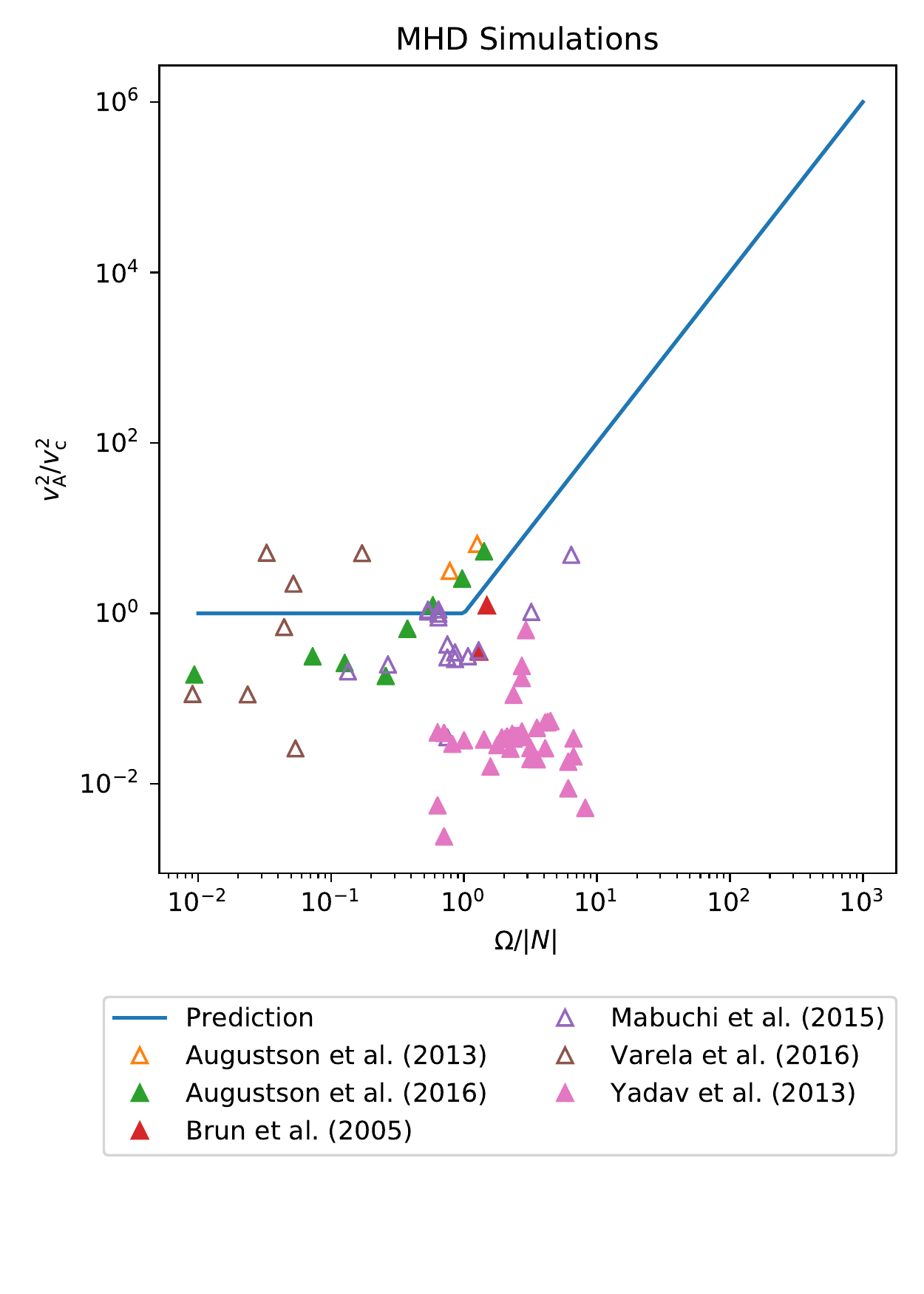}
\caption{The ratio $\varv_{\rm A}^2 / \varv_{\rm c}^2$ of magnetic energy to convective kinetic energy is shown alongside our predictions as a function of $\Omega/|N|$ for a variety of MHD convection simulations. Filled shapes indicate simulations performed on spherical domains while open shapes indicate those performed in shellular domaints. All of these data were averaged over the simulation domain. Our prediction is constant for slow rotation ($\Omega < |N|$) and scales as $\Omega/|N| \approx (\Omega/|N|)^{2}$ for rapid rotation ($\Omega > |N|$).}
\label{fig:e_mag}
\end{figure}

Finally we examine the ratio of magnetic to kinetic energy, shown as a function of $\Omega/|N|$ in Fig.~\ref{fig:e_mag}.
Our predictions are shown as a solid line and the results of simulations are shown as a variety of shapes, following the previous convention that solid shapes indicate spherical domains, open shapes indicate shellular domains, and triangles denote measurements averaged over the simulation domain.
Note that our predictions for this energy ratio are not new: they are identical to those made by several authors including~\citet{2002Icar..157..426S} and~\citet{2016ApJ...829...92A}.

There is significant scatter between simulations performed by different groups but within the sets of simulations with the most dynamic range in $\Omega/|N|$ we see a plateau towards slow rotation and scaling similar to that of our predictions for rapid rotation.

Our predicted scalings are also similar to those that found by several groups in fits to their own numerical simulations.
For instance at fixed heat flux we predict that the magnetic field strength scales as  $B \propto \varv_{\rm A} \propto \Omega^{1/4}$ and in simulations this scaling has been found to be $\Omega^{-0.02}$~\citep{doi:10.1111/j.1365-246X.2006.03009.x}\footnote{See their equation (33). Note that $\mathrm{Lo} \propto \varv_{\rm A}/\Omega$ and their $\mathrm{Ra}_Q^* \propto \Omega^{-3}$, such that $\mathrm{Lo} \propto (\mathrm{Ra}_Q^*)^(0.34)$ means that $\varv_{\rm A} \propto \Omega^{1-0.34\times 3}$.}, $\Omega^{-0.11}$~\citep{2013Icar..225..185Y}\footnote{From their dipolar fit with no magnetic Prandtl number dependence.}, $\Omega^{0}$~\citep{2013Icar..225..185Y}\footnote{From their dipolar fit with magnetic Prandtl number dependence.}, and between $\Omega^{-0.02}$ and $\Omega^{0.19}$~\citep{2017JFM...813..558A}\footnote{See their figure 11(a). Their $\epsilon$ is proportional to $\Omega^{-3}$ and their $\lambda \propto B \Omega^{-1}$.}.

%%%---------- close: mhd_sim_tests

%%%%%%%%% jump to xi_sims
%%%---------- open: xi_sims

\subsection{Baroclinicity}
\label{sec:xi_sims}

\begin{figure}
\centering
\includegraphics[width=0.47\textwidth]{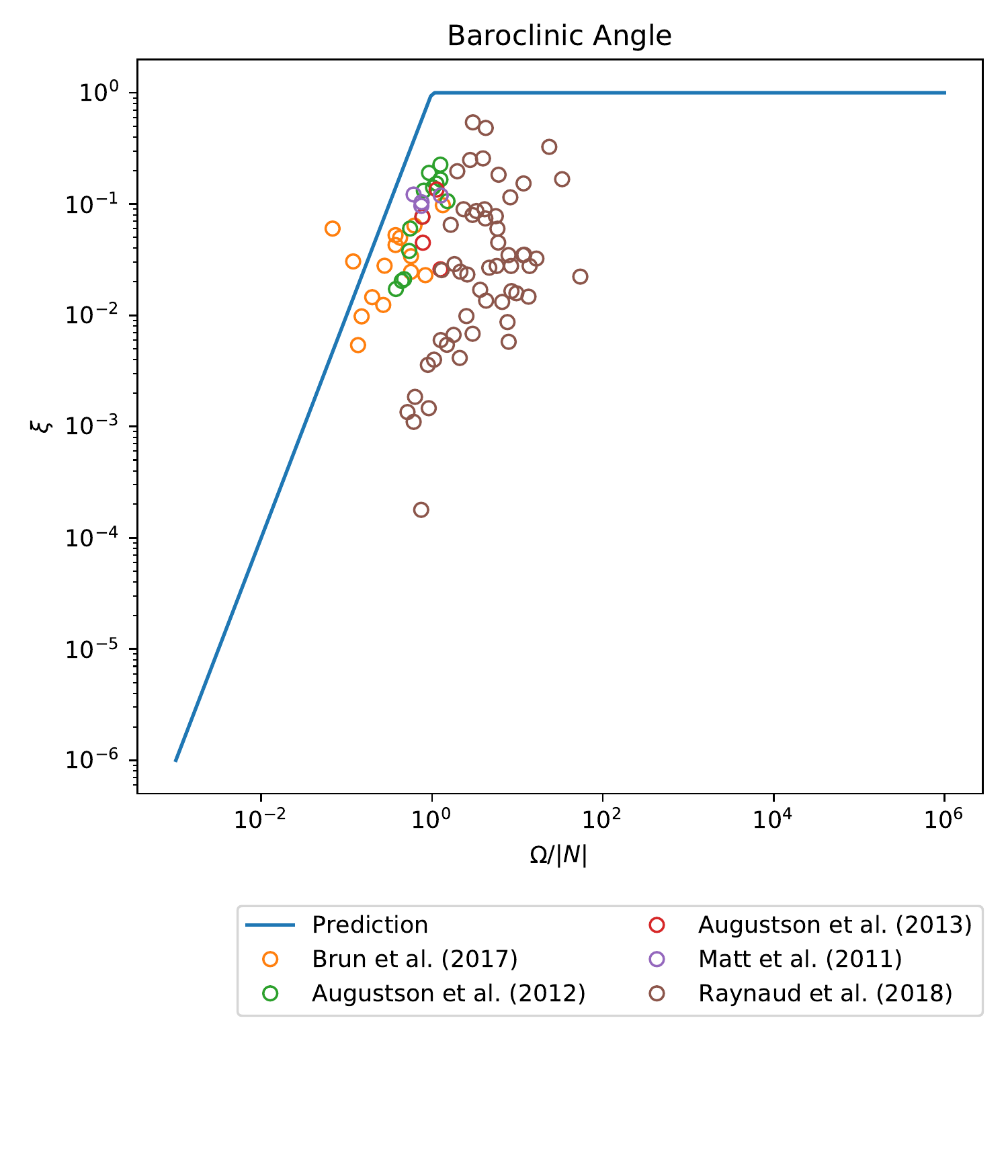}
\caption{The angle $\xi$ between the pressure gradient and the entropy gradient is shown for a variety of simulations. Our prediction scales as $(\Omega/|N|_0)^2$ for slow rotation ($\Omega < |N|_0$) and is constant for rapid rotation ($\Omega > |N|_0$).}
\label{fig:xi_sims}
\end{figure}

As described in appendix~\ref{appen:data}, we have inferred $\xi$ from the reported results of various hydrodynamic and MHD simulations.
Our prediction is the same for both kinds of simulations and so we analyze them together.
Some authors reported the temperature difference between different latitudes in a model while others reported latitudinal gradients of closely related quantities like the Nusselt number.
The results of our inference and standardization are shown in Fig.~\ref{fig:xi_sims}.

We find reasonable agreement with our predicted $\Omega^2$ scaling for slow rotators and, as expected, we see $\xi$ saturating in the limit of rapid rotation.
There is considerable scatter among simulations in the saturated regime and even considerable disagreement amongst simulations performed by the same group with the same code.
This may reflect a sensitivity of the saturated $\xi$ to the geometry or other dimensionless parameters in the simulation.

These data support our conclusion in Part~I that the thermal wind term is one of the dominant terms in both the hydrodynamic and MHD regimes and in the limits of both slow and rapid rotation, in agreement with the arguments of~\citet{2012MNRAS.426.1546B}.
The scaling we find suggests that the baroclinic angle in convection zones is indeed driven by rotation-induced anisotropy in the convective heat flux~\citep{2018MNRAS.480.5427J}.%%%---------- close: xi_sims
%%%---------- close: xi_sims

%%%%%%%%%%% jump to future
%%%---------- open: future
\section{Future Tests}
\label{sec:future}

We have tested our predictions under many different circumstances but there are further tests that would be fruitful in the future.
First, while we have tested our predictions for the latitudinal and spherical radial shear, we have not made a quantitative test of our prediction that the cylindrical vertical shear $\partial_z\Omega$ is suppressed relative to the radial component.
This is seen in simulations~\citep[see e.g.][]{2015ApJ...806...10M} but so far as we are aware it has not been quantified.
It ought to be straightforward to add $\partial_R\Omega$ and $\partial_z\Omega$ to the output of future simulations or even to run previous simulations again with these additional outputs.

Secondly, it would be good to test our predictions for the meridional circulation velocities $u_r$ and $u_\theta$.
There have been measurements of the meridional circulation in the Sun~\citep{2013ApJ...774L..29Z,2015ApJ...813..114R,2020ApJ...890...32S} but there are still significant disagreements between the different inversion techniques which make a direct comparison to our theory challenging.
On the other hand, such a comparison could be done with simulations by reporting the time- and azimuthal-average of the meridional velocity field.
As far as we are aware this has not been reported~\footnote{Some authors~\citep[e.g.][]{2012ApJ...756..169A} report the root-mean square of the meridional circulation. Despite the similar name this is not the quantity we predict. By squaring before taking a time-average they are measuring the strength of axisymmetric convective motions rather than the slow mean meridional flow.} but we see no particular barrier to doing so.

More broadly, further tests from simulations primarily face the challenge of reaching high enough P\'{e}clet numbers to be applicable to efficient convection.
The primary limitation to this is the availability of computing resources, so we hope that more simulations will be performed at $\mathrm{Pe} \approx 10^{3}$ to $10^{5}$ in the near future.

Finally, the most useful observational comparisons are those that probe the most rapidly-rotating systems.
These are the systems which are best able to test the slopes we have predicted and to thereby either verify or falsify our theories.
In this regard data such as the differential rotation measurements of~\citet{2015ApJ...806..212D} are extremely useful because, while they report observations of a single system, it is well-characterized and extremely rapidly rotating, particularly relative to its slow convective turnover time.
This allows it to more strongly constrain the scaling of differential rotation.%%%---------- close: future

%%%%%%%%%%% jump to discussion
%%%---------- open: discussion
\section{Prior Predictions}

In Part~I we made predictions for the scaling of the differential rotation, meridional circulation, magnetic field and baroclinic angle.
We are aware that two of our predicted scalings have been suggested previously.
These are that for the magnetic field~\citep{2002Icar..157..426S} and that of convection speed with rotation rate the hydrodynamic limit~\citep{1979GApFD..12..139S}.
The remaining scalings are new.

However, there have been many previous predictions of the same quantities and it is worth comparing our predictions with these.
Our prediction that $|R\nabla\ln\Omega|$ is a constant of order unity for slowly-rotating systems is consistent with predictions based on the $\Lambda$ effect~\citep[e.g.][and references therein]{2019A&A...622A..40K}.
\citet{1979ApJ...229.1179G} predicted that for slowly-rotating systems the angular momentum of fluid parcels is conserved in convection and so that $\Omega \propto r^{-2}$, or $|R\nabla\ln\Omega| = 2$, consistent with our predictions.
\citet{2015ApJ...808...35K} make the same prediction of $\Omega \propto r^{-2}$ for slowly-rotating systems and provide two possible scalings for rapidly-rotating ones, $|R\nabla\ln\Omega| = \alpha, 1 < \alpha < 3/2$ and $\Omega\approx \Omega_0 + |N|$ for some suitable choice of $\Omega_0$.
Both of these are inconsistent with our predictions, particularly the latter, which introduces a new characteristic frequency-scale $\Omega_0$ and predicts a characteristic length-scale for the shear of $(\partial_r \ln |N|)^{-1} \approx h$.

Our prediction for the scaling of the convection speed ($\varv_{\rm c} \approx h |N|^{2} \Omega^{-1}$) is based on the theory of~\citet{1979GApFD..12..139S}, and has been suggested subsequently along various different lines by~\citet{2002Icar..157..426S},~\citet{2010SSRv..152..565C} and~\citet{2014ApJ...791...13B}.
In the hydrodynamic case, our prediction for the scaling of $|N|$ with $\Omega$ matches that of~\citet{2014ApJ...791...13B} as well.
We previously predicted a somewhat different scaling~\citep{jermyn} of $\varv_{\rm c} \approx h |N|^{3/2} \Omega^{-1/2}$ because, unlike~\citet{1979GApFD..12..139S},~\citet{jermyn} did not impose the lower bound on the vertical wavenumber.
Such a lower bound is physically motivated for stars by the finite scale height, so we favour the scaling of~\citet{1979GApFD..12..139S} here.

Our prediction for the scaling of the the magnetic field with rotation rate ($\varv_{\rm A} \approx h |N| \approx h |N|_0^{3/4} \Omega^{1/4}$) is based on the relationship between heat flux and magnetic field by~\citet{doi:10.1111/j.1365-246X.2006.03009.x} and the bound that if $\varv_{\rm A} \ga h |N|$ then the magnetic field begins to inhibit the linear convective instability~\citep{1966MNRAS.133...85G}.
This prediction agrees with the theory of~\citet{1979GApFD..12..139S}, and~\citet{2002Icar..157..426S} predict both the same scaling for $\varv_{\rm A}$ and for $|N|$.
However, a variety of other scaling laws have been proposed.
These are summarized in Section~\ref{sec:mhd}.
They generally agree that, at fixed $|N|$, the magnetic field strength varies as a small power of $\Omega$ ranging from $-0.02$ to $0.19$, all of which is somewhat smaller than we have predicted.

\section{Discussion}
\label{sec:discussion}

Up to this point we have focused on testing our predictions against observations and simulations.
We have found general agreement on the existence and location of plateaus, as well as on the trends away from plateaus.
Importantly, in section~\ref{sec:sims}, we found that simulations show a strong dependence on P\'{e}clet number and that they only converge to our predictions as $\mathrm{Pe}$ becomes large in the limit of efficient convection.
Because this is the most common limit in astrophysical systems it matters that simulations attempting to reproduce observed differential rotation operate in this regime.

We now turn to exploring broader astrophysical implications.
We have predicted that the relative shear is of order unity for slowly rotating convection zones and falls off quickly once the rotation becomes faster than the convective turnover frequency.
In particular, the relative shear per unit $\log r$ is at most of order $|N|/\Omega$ in the rapidly-rotating limit.
A consequence of this is that the shear is approximately bounded by the \brvs\ frequency, which is of order the convective turnover frequency.
This is because if $\Omega > |N|$ then the absolute shear per unit $\log r$ is $|N|$ while if $\Omega < |N|$ it is $\Omega$, which is less than \brvs\ frequency $|N|$.
Hence we expect the maximum absolute shear across a convection zone to be
\begin{align}
	\Delta\Omega_{\rm max} \approx \int_{\rm convection\,\, zone} |N| d\ln r.
	\label{eq:d_omega_max}
\end{align}
Note that this upper bound is robust even if, as suggested by Fig.~\ref{fig:mhd}, the differential rotation falls off faster with $\Omega$ than we have predicted.

Because, in the MHD regime, $|N|$ deviates only weakly from $|N|_0$ we may estimate the right-hand side of equation~\eqref{eq:d_omega_max} using $|N|_0$ from stellar models.
For the red giants studied by~\citet{2015A&A...580A..96D} $\Delta\Omega_{\rm max} \approx 4\times 10^{-6}\,{\rm s}^{-1}$.
This is on the order of what ~\citet{2014ApJ...788...93C} and~\citet{0004-637X-808-1-35} suggest is needed to match asteroseismic inferences of core-envelope shear in red giants, so an order unity pre-factor in equation~\eqref{eq:d_omega_max} would suffice to bring our bound into agreement with their calculations, and it seems possible that most or all of the differential rotation in such stars is in their convective envelopes.
However, because equation~\eqref{eq:d_omega_max} is an upper bound, it remains possible that a substantial component lies in the radiative zone~\citep{2019MNRAS.485.3661F} and some astereoseismic evidence indicates that there is more shear in the radiative zones of red giants than in their convection zones~\citep{Di_Mauro_2018}.

A further prediction is that rapidly-rotating convecting stars ought to exhibit significant equator to pole temperature differences.
These differences are due to the Coriolis effect acting on convective motions and not the von~Zeipel effect.
This has been seen in simulations~\citep{2018A&A...609A.124R} and observations~\citep{2011ApJ...732...68C}, and our scaling relations complement these by providing a theoretical curve with which to interpret the observations.
In particular, in the usual notation,
\begin{align}
	T_{\rm eff} \propto g_{\rm eff}^{\beta},
\end{align}
where $g_{\rm eff}$ is the effective acceleration due to gravity and centrifugal effects and $\beta=1/4$ is the von~Zeipel exponent~\citep{1924MNRAS..84..665V}.
Differentiating this with respect to latitude yields
\begin{align}
	\frac{\partial \ln T_{\rm eff}}{\partial \theta} = \beta \frac{\partial \ln g_{\rm eff}}{\partial \theta} \approx \beta \frac{\Omega^2 r \cos\theta}{g}.
\end{align}
Our findings (equation~\ref{eq:xi_solar}) suggest instead that
\begin{align}
	\frac{\partial \ln T_{\rm eff}}{\partial \theta} = \xi \frac{r |N|^2}{g_{\rm eff}}.
\end{align}
For slowly-rotating systems, $\xi \propto \Omega^2/|N|^2$ and this becomes
\begin{align}
	\frac{\partial \ln T_{\rm eff}}{\partial \theta} \propto \frac{\Omega^2 r}{g_{\rm eff}},
\end{align}
which is equivalent to the von~Zeipel result, possibly with a different $\beta$.
For rapidly-rotating systems with $\Omega > |N|$, $\xi$ saturates and
\begin{align}
	\frac{\partial \ln T_{\rm eff}}{\partial \theta} \approx \frac{r |N|^2}{g_{\rm eff}}.
\end{align}
This saturation could be why observations of late-type stars have inferred $\beta$ smaller than $1/4$: if the systems enter the saturated regime then a fit to the unsaturated scaling would tend to estimate a lower $\beta$~\citep{2011ApJ...732...68C}.

For extremely rapid rotators we also predict that convection is suppressed.
In stars this leads to an increase in the \brvs\ frequency $N$ and thence a steepening of the entropy gradient to carry the nuclear luminosity.
This is likely to be difficult to observe directly in measurements such as the stellar radius or effective temperature because the effect is even smaller than centrifugal expansion (see appendix~\ref{appen:expansion}).
On the other hand, it could be observed in asteroseismic measurements.
The cores of massive stars likely emit internal gravity waves, with a peak frequency thought to be proportional to the \brvs\ frequency~\citep{Rogers_2013,couston_lecoanet_favier_le_bars_2018}.
Moreover convection in these stars is so slow that nearly all of them are in the rapidly rotating limit.
Hence if our predictions hold then there ought to be a systematic shift in the peak frequency of these waves as a function of rotation that scales as $\Omega^{1/4}$.

Finally, note that in the very rapid limit we expect the Taylor-Proudman constraint to hold.
That is we predict that the star forms nested cylinders of constant rotation because $|\partial_z \Omega| \ll |\nabla \Omega|$.
We could not find data to probe this prediction quantitatively but essentially all of the simulations we examined show this feature in the rapidly rotating limit.%%%---------- close: discussion

\section*{Acknowledgements}

%%%%%%%%%%% jump to Snippets/acknowledgements
%%%---------- open: Snippets/acknowledgements
The Flatiron Institute is supported by the Simons Foundation. ASJ thanks the Gordon and Betty Moore Foundation (Grant GBMF7392) and the National Science Foundation (Grant No. NSF PHY-1748958) for supporting this work.
%%%---------- close: Snippets/acknowledgements
%%%%%%%%%%% jump to acknowledgements
%%%---------- open: acknowledgements
ASJ also acknowledges financial support from a UK Marshall Scholarship as well as from the IOA, ENS and CEBS to work at ENS Paris and CEBS in Mumbai.
PL acknowledges travel support from the french PNPS (Programme National de Physique Stellaire) and from CEBS.
CAT thanks Churchill College for his fellowship.
SMC is grateful to the IOA for support and hospitality and thanks the Cambridge-Hamied exchange program for financial support.
ASJ, SMC and PL thank Bhooshan Paradkar for productive conversations related to this work.
The authors thank Jim Fuller, Douglas Gough, Steven Balbus, Chris Thompson for comments on this manuscript.
We are also thankful to the anonymous referee, whose suggestions led us to include more data as well as the sensitivity analysis in Appendices~\ref{appen:Navg} and~\ref{appen:models}.
ASJ is grateful to Lars Bildsten for helpful suggestions regarding the data analysis, to Frank Timmes and Matteo Cantiello for suggestions regarding the presentation, and to Daniel Lecoanet, Evan Anders and Matthew Browning for comments regarding the interpretation of simulations.%%%---------- close: acknowledgements

\section*{Data Availability}

Details of how the data from various sources were obtained, standardized and processed are provided in Appendix~\ref{appen:data}, and all scripts and files needed to reproduce this analysis are available on \url{https://doi.org/10.5281/zenodo.3992227}.

\bibliographystyle{mnras}
\bibliography{refs}

\appendix

%%%%%%%%%%% jump to data_processing
%%%---------- open: data_processing
\section{Data Processing}
\label{appen:data}

Here we explain the way in which data were processed to produce our figures.
All scripts and inlists as well as auxiliary data files used in this data processing are available on \url{https://doi.org/10.5281/zenodo.3992227}.

In many cases the data processing required stellar models.
These were produced using revision 11701 of the Modules for Experiments in Stellar Astrophysics
\code{MESA}.
The \code{MESA} equation of state (EOS) is a blend of the OPAL~\citep{Rogers2002}, SCVH~\citep{Saumon1995}, PTEH~\citep{Pols1995}, HELM~\citep{Timmes2000} and PC~\citep{Potekhin2010} equations of state.
Radiative opacities are primarily from OPAL~\citep{Iglesias1993,
Iglesias1996}, with low-temperature data according to~\citet{Ferguson2005}
and the high-temperature, Compton-scattering dominated regime by~\citet{Buchler1976}.
Electron conduction opacities are according to~\citet{Cassisi2007}.
Nuclear reaction rates are from JINA REACLIB~\citep{Cyburt2010} plus additional
tabulated weak reaction rates~\citep{Fuller1985, Oda1994, Langanke2000}.
Electron screening is included via the prescription of~\citet{Chugunov2007}.
Thermal neutrino loss rates are according to~\citet{Itoh1996}.
Mixing length theory was implemented following~\citet{1968pss..book.....C} with mixing length parameter $\alpha = 2$.
Models were created on the pre-main sequence and evolved from there.
All other parameters were set to their defaults.

\subsection{Latitudinal Shear}

We describe the observations.

\subsubsection{Benomar Observations}

Data were taken from tables~S2 and~S3 of~\citet{2018Sci...361.1231B}, who inferred differential rotation through asteroseismology  for several solar-type stars and reported it in the form of the latitudinal shear and mean rotation rates.
The reported shear $\Omega_1$ is $2/15$ times the difference between the pole and the equator.
There are $\pi/2$ radians between the pole and the equator so we approximate
\begin{align}
	|R\nabla\Omega| \approx |\partial_\theta \Omega| \approx \frac{2}{\pi}\left(\frac{15}{2}\right)\Omega_1.
\end{align}

Uncertainties in both the shear and mean rotation rate were reported.
These were propagated into the relative shear by the formula
\begin{align}
	d\mu = \sqrt{\sum_j \left(\frac{\partial \mu}{\partial x_j}\right)^2 dx_j^2},
	\label{eq:dmu}
\end{align}
where $\mu$ is a function of interest and $x_j$ are variables on which it depends.
This amounts to the assumption that errors are uncorrelated and small relative to the scale over which higher-order derivatives are relevant.

Stellar models were constructed which matched the ages, metallicities and masses provided in their table~S2.
The average $|N|_0$ (\brvs\ frequency) in the convection zone of each model was calculated following equation~\eqref{eq:N_avg}.

\subsubsection{Bazot Observations}

Data were taken from tables~1 and~A3 of~\citet{2019A&A...623A.125B} who inferred differential rotation through asteroseismology  for several solar-type stars and reported in the form of the latitudinal shear and mean rotation rates.
The reported shear $\Omega_1$ is $2/15$ times the difference between the pole and the equator.
There are $\pi/2$ radians between the pole and the equator so we approximate
\begin{align}
	|R\nabla\Omega| \approx |\partial_\theta \Omega| \approx \frac{2}{\pi}\left(\frac{15}{2}\right)\Omega_1.
\end{align}

Uncertainties in both the shear and mean rotation rate were reported for 16~Cyg~A.
For 16~Cyg~B a weighted average of the three modes was used following the recommendation in the caption of table~1.
The uncertainty of this average was computed as the weighted root-mean-squared difference between the three modes and the average.

Uncertainties were propagated into the relative shear equation~\eqref{eq:dmu}.
Stellar models were constructed which matched the ages, metallicities and masses provided in their table~A3, and the average $|N|_0$ in the convection zone of each model was calculated according to equation~\eqref{eq:N_avg}.

\subsubsection{Ammler-von Eiff Observations}
\label{appen:ammler}

Data were taken from table~2 of~\citet{2012A&A...542A.116A}.
The latitudinal differential rotation was spectroscopically inferred for several A to F stars and reported in the form of the coefficient $\eta/\sqrt{\sin i}$, where $\eta$ is the pole to equator shear divided by the equatorial angular velocity.
For each object our hypothesis is that the latitudinal differential rotation arises owing to the surface convection zone.
So we identify
\begin{align}
	|R\nabla\Omega| \approx |\partial_\theta \Omega| \approx \frac{2}{\pi}|\Omega_{\rm pole}-\Omega_{\rm equator}|.
	\label{eq:eta_shear}
\end{align}

Stars without a surface convection zone were filtered out of the sample.
Many systems were consistent with zero shear to within the error bars.
So we only include those with at least a $1\sigma$ detection of shear.

The rotation rate is reported as $\varv \sin i$, where $\varv$ is the surface rotation velocity.
Inclinations were not reported.
So rotation speeds are considered to be lower bounds and shear measurements to be upper bounds.

Eighty stellar models were constructed with metallicity $Z=0.02$ and spaced uniformly in mass from $0.2\, M_\odot$ to $1.4\, M_\odot$.
The models were evolved to an age of $(M/M_\odot)^{-2.5}{\rm GYr}$ or $10\,{\rm GYr}$, whichever was smaller.
For each object the model which best-matched the reported $T_{\rm eff}$ in their table~5 was used to compute $\Omega = \varv/R$ and $|N|_{\rm avg}$.

\subsubsection{Frasca Observations}
\label{appen:frasca}

Data were taken from~\citet{2011A&A...532A..81F}.
The differential rotation was inferred from star spot observations of the K star KIC~8429280 and reported in the form of the pole to equator shear, which we treat with equation~\eqref{eq:eta_shear}.
The rotation rate is reported as $\varv \sin i$, where $\varv$ is the surface rotation velocity.
The inclination is also reported and so we use it to compute $\varv$.
We employ the same grid of main-sequence stellar models from appendix~\ref{appen:ammler} to compute $|N|_{\rm avg}$ and $R$ and use these to calculate $\Omega = \varv/R$ and $\Omega/|N|_{\rm avg}$.

\subsubsection{Bonanno Observations}

Data were taken from~\citet{2014A&A...569A.113B} who report differential rotation inferred from star spot observations of the late F star KIC~5955122.
The differential rotation was reported in the form of a pole to equator shear, which we treat with equation~\eqref{eq:eta_shear}.
We normalise this to the mean rotation rate and take the result to be approximately $|R\nabla\Omega|/\Omega$.
A stellar model was constructed which matched the age, metallicity and mass reported for KIC~5955122 and we used this to compute $|N|_{\rm avg}$ and $\Omega/|N|_{\rm avg}$.

\subsubsection{Donati Observations}
\label{appen:donati}

Data were taken from~\citet{2008MNRAS.390..545D}.
The differential rotation was inferred for a sample of M-dwarfs from spectropolarimetry and reported as the surface equator to pole difference in $\Omega$ normalised against that of the equator.
We treat this with equation~\eqref{eq:eta_shear}.

\citet{2008MNRAS.390..545D} report the Rossby number in the form of $\tau_{\rm c} P_{\rm rotation}$, where $\tau_{\rm c} = 1 / |N|_0$~\citep{1980LNP...114...19G} was determined from an empirical calibration to non-rotating stellar models~\citep{2007AcA....57..149K} and $P_{\rm rotation}$ is the average spin period of the star.
So we take $\frac{\Omega}{|N|_0} \approx 2\pi \mathrm{Ro}^{-1}$.

\subsubsection{Reinhold Observations}

Data were taken from~\citet{2015A&A...583A..65R}, who report differential rotation inferred from star spot observations of \emph{Kepler} main-sequence and subgiant stars.
Stars which were flagged as having highly stable periods were excluded as possible binary or pulsator systems.

The differential rotation was reported in the form of the coefficient $\eta/\sqrt{\sin i}$, which we handle as in appendix~\ref{appen:frasca}.
We employ the same grid of main-sequence stellar models from appendix~\ref{appen:ammler} to compute $|N|_{\rm avg}$ and $R$ and use these to calculate $\Omega = \varv/R$ and $\Omega/|N|_{\rm avg}$.
Objects hotter than the hottest model in the grid or cooler than the coolest model in the grid were excluded from the sample.

\subsubsection{Lurie Observations}

Data were taken from~\citet{2017AJ....154..250L}, who report differential rotation inferred from star spot observations of \emph{Kepler} eclipsing binary systems.
They infer the parameter $\eta$ by
\begin{align}
	\eta = \frac{P_{\rm max}-P_{\rm min}}{P_{\rm max}},
	\label{eq:eta}
\end{align}
where the periods $P_{\rm min}$ and $P_{\rm max}$ are the first minimum and maximum in the periodogram of the lightcurve of the system.
They report these periods and so we compute $\eta$ using equation~\eqref{eq:eta} and translate this into a shear using equation~\eqref{eq:eta_shear}.

Triple-star systems were excluded by means of the catalogues of~\citet{2013ApJ...768...33R} and~\citet{2016MNRAS.455.4136B} on the grounds that the inference of~\citet{2017AJ....154..250L} is based on the assumption of a binary system.
We further exclude the false positive systems described by~\citet{2016MNRAS.455.4136B} because these have unusual light curves which could interfere with the inference of rotation periods.

Objects were matched by effective temperature to the grid of main-sequence stellar models from appendix~\ref{appen:ammler}, from which we computed $|N|_{\rm avg}$ and $\Omega/|N|_{\rm avg}$.
The effective temperatures of these objects were obtained from the Kepler Eclipsing Binary catalog~\citep{2012AJ....143..123M,2014PASP..126..914C,2015MNRAS.452.3561L,2016AJ....151..101A}.
Objects with no temperature in the catalogue were excluded.
Objects hotter than the hottest model in the grid or cooler than the coolest model in the grid were excluded from the sample.

\subsubsection{Davenport Observations}

Data were taken from~\citet{2015ApJ...806..212D}, who report differential rotation inferred from star spot observations of the M-dwarf GJ~1243.
The spots reveal a shear of $0.0047\,{\rm rad\,d^{-1}}$.
After accounting for spot latitude, they report a pole-equator shear of $0.012\pm 0.002\,{\rm rad\, d^{-1}}$.
The rotation period of the planet is reported as $0.592596\pm 0.00021\,{\rm d}$.
We convert this into $\eta$ and then use equation~\eqref{eq:eta_shear} to obtain a relative shear.
Uncertainties were propagated along the way with equation~\eqref{eq:dmu}.
The Rossby number is also reported, as by~\citet{2008MNRAS.390..545D} and we handle it in the same manner as in appendix~\ref{appen:donati}.

\subsection{Radial Shear}

We describe the observations.

\subsubsection{Deheuvels Observations}

Data were taken from~\citet{2015A&A...580A..96D}.
The differential rotation was seismically inferred for several red giants and reported in the form of the difference between the core and envelope rotation rates divided by the envelope rotation rate.
We take this to equal
\begin{align}
	\ln \frac{\Omega_{\rm core}}{\Omega_{\rm envelope}} \approx \int_{\rm convection\ zone} |R\nabla \ln \Omega| d\ln R,
\end{align}
on the assumption that most of the differential rotation is developed in the convection zone.
We thereby infer
\begin{align}
	|R\nabla\ln\Omega|_{\rm avg} \approx \ln \frac{\Omega_{\rm core}}{\Omega_{\rm envelope}}\left(\int_{\rm convection\ zone} d\ln R\right)^{-1}.
\end{align}

Stellar models were constructed with \code{MESA} which matched the masses and radii in table~2 of~\citet{2015A&A...580A..96D}.
These were used to compute $|N|_{\rm avg}$ and $\int_{\rm convection\ zone} d\ln R$.

\subsubsection{Klion Observations}

Data were taken from the text of~\citet{doi:10.1093/mnrasl/slw171} for the red giant Kepler-56.
The differential rotation was seismically inferred and reported in the form of a power-law
\begin{align}
	\Omega(r) \propto r^\beta
\end{align}
with $\beta \approx 1$, so we take $|R\nabla\Omega|/\Omega \approx 1$.
The rotation period is provided in the text as $74\pm 3{\rm d}$.
A solar-metallicity stellar model was constructed with \code{MESA} which matched the mass and radius reported for Kepler~56.
This was used to compute $|N|_{\rm avg}$.

\subsubsection{Nielsen Observations}

Data were taken from table~1 of~\citet{2017A&A...603A...6N}.
The differential rotation was seismically inferred for several solar-like stars and reported as envelope and interior angular frequencies.
We then approximate the shear as
\begin{align}
	|R\nabla\Omega| \approx R\partial_R \Omega \approx \frac{1}{\Delta \ln R}\left(\frac{\max(\Omega_{\rm core},\Omega_{\rm envelope})}{\min(\Omega_{\rm core},\Omega_{\rm envelope})} - 1\right),
	\label{eq:core_env}
\end{align}
where $\Delta \ln R$ was computed as
\begin{align}
	\Delta \ln R = \int_{\rm convection\,\, zone} \frac{dr}{\max(r,h)}
\end{align}
as we have done in equation~\eqref{eq:N_avg}.
Uncertainties were propagated via equation~\eqref{eq:dmu}.
Objects were matched by effective temperature to the grid of main-sequence stellar models from appendix~\ref{appen:ammler}, from which we computed $|N|_{\rm avg}$ and $\Omega/|N|_{\rm avg}$.

%%%%%%%%% jump to hydro_sims_appen
%%%---------- open: hydro_sims_appen

\subsection{Hydrodynamic Simulations}

We describe how we extract the relevant data from various hydrodynamic simulations.

\subsubsection{Browning 2004 Simulations}
\label{appen:browning04}

These data come from three-dimensional anelastic hydrodynamic simulations of convection in a rotating spherical domain~\citep{2004ApJ...601..512B}.
The differential rotation is reported as the difference in angular velocity between the equator and a latitude of $\pi/3$, normalized to the equatorial angular velocity, so we write
\begin{align}
	|R\nabla \ln \Omega| \approx \frac{3}{\pi} \frac{\Omega_{\rm equator}-\Omega_{\pi/3}}{\Omega_{\rm equator}}.
	\label{eq:0_60}
\end{align}

\citet{2004ApJ...601..512B} report the Reynolds number
\begin{align}
	\mathrm{Re} = \frac{\tilde{\varv} L}{\nu},
\end{align}
where $\tilde{\varv}$ is the root-mean square of the convective velocity, $L$ is the radial extent of the convective region and $\nu$ is the microscopic viscosity.
They also report the Prandtl number
\begin{align}
	\mathrm{Pr} = \frac{\nu}{\alpha} = 0.25
\end{align}
and the Rayleigh number
\begin{align}
	\mathrm{Ra} = \frac{g \Delta s L^3}{\rho \nu \alpha}\frac{\partial \rho}{\partial s},
	\label{eq:Rayleigh_Nrho}
\end{align}
where $\alpha$ is the thermal diffusivity, $\rho$ is the density, $s$ is the entropy, $\Delta s$ is the entropy change across the domain and $g$ is the acceleration owing to gravity.
We assume that the partial derivative is taken at constant pressure,
\begin{align}
	\left(\frac{\partial \ln \rho}{\partial s}\right)_P = \frac{\gamma-1}{\gamma}.
\end{align}
With this information we obtain the convective velocity by
\begin{align}
	\frac{\varv_{\rm c}}{h |N|} \approx \mathrm{Re} \sqrt{\frac{(\gamma-1) \mathrm{Pr}}{\gamma \mathrm{Ra}}}
	\label{eq:vc_hN}
\end{align}
and the P\'{e}clet number as
\begin{align}
	\mathrm{Pe} = \mathrm{Re} \mathrm{Pr} \approx \sqrt{\mathrm{Ra} \mathrm{Pr}}.
	\label{eq:Pe}
\end{align}

The Rossby number actually realised in the flow is also reported.
This they define as
\begin{align}
	\mathrm{Ro} = \frac{\omega_{\phi,\mathrm{convective}}}{2 \Omega},
	\label{eq:vortRo}
\end{align}
where $\omega_{\phi,\mathrm{convective}} \approx \varv_{\rm c}/h$ is the root-mean square of the vorticity of the convective flow.
So we find
\begin{align}
	\frac{\Omega}{|N|} \approx \frac{1}{2\mathrm{Ro}}\left(\frac{\varv_{\rm c}}{h |N|}\right),
	\label{eq:OmegaN_aug}
\end{align} 
where the term in parentheses is given by equation~\eqref{eq:vc_hN}.

\subsubsection{Augustson 2012 Simulations}
\label{appen:aug12}

These data come from three-dimensional anelastic hydrodynamic simulations of convection in a rotating spherical shell domain~\citep{2012ApJ...756..169A}.
The differential rotation is reported as the difference in angular velocity between the equator and a latitude of $\pi/3$, normalized to the equatorial angular velocity, so we process this as equation~\eqref{eq:0_60}.
The Rossby number actually realised in the flow is also reported, as equation~\eqref{eq:vortRo}, so we handle this as before.

The temperature difference $\Delta T$ between the equator and a latitude of $\pi/3$, averaged over depth, is reported.
To obtain the baroclinic angle $\xi$ from $\Delta T$ we write
\begin{align}
	\xi &\approx \frac{\partial_\theta s}{r \partial_r s}.
	\label{eq:xi_approx}
\end{align}
This is valid in the limit of a radial pressure gradient and a mostly-radial entropy gradient, so that $\xi$ is the small projection of $\nabla s/|\nabla s|$ along $e_\theta$.
\citet{2012ApJ...756..169A} define
\begin{align}
	{\rm Ra} = \frac{\Delta S g L^3}{c_p \nu \alpha},
\end{align}
where ${\rm Ra}$ is the Rayleigh number, $\Delta S$ is the radial change in dimensionful entropy across the simulation domain, $L$ the vertical size of the domain, $g$ is the acceleration owing to gravity, $c_{P}$ is the specific heat at constant pressure, $\nu$ is the viscosity and $\alpha$ is the thermal diffusivity.

The ratio $\Delta S / c_{P}$ is related to the dimensionless entropy change $\Delta s$ by
\begin{align}
	\Delta s = \frac{\Delta S}{c_{P}} \left(\frac{\gamma}{\gamma-1}\right).
	\label{eq:s_dim}
\end{align}
Identifying
\begin{align}
	\Delta s \approx L \partial_r s,
\end{align}
we then obtain
\begin{align}
	r \partial_r s \approx {\rm Ra} \frac{\nu \alpha}{g L^4}\left(\frac{\gamma}{\gamma-1}\right).
\end{align}
They further let the Prandtl number $\rm Pr$ be $\nu / \alpha$ and the Taylor number $\rm Ta$ be $4\Omega^2 L^4 / \nu^2$.
With these we obtain
\begin{align}
	r\partial_r s \approx 4\frac{\gamma}{\gamma-1}\left(\frac{\rm Ra}{\rm Pr\,Ta}\right)\left(\frac{\Omega^2 r}{g}\right).
\end{align}
So
\begin{align}
	\xi \approx \frac{\gamma-1}{4\gamma}\left(\frac{\rm Pr\,Ta}{\rm Ra}\right)\left(\frac{g}{\Omega^2 r}\right)\partial_\theta s.
\end{align}
Assuming that isobars are nearly spherical, we see that
\begin{align}
	\partial_\theta s \approx \partial_\theta \ln T \left.\frac{\partial s}{\partial \ln T}\right|_P.
	\label{eq:dsdlnT}
\end{align}
With equation~\eqref{eq:s} we find
\begin{align}
	\left.\frac{\partial s}{\partial \ln T}\right|_P = \frac{\gamma}{\gamma-1},
\end{align}
so
\begin{align}
	\xi \approx \frac{1}{4}\left(\frac{\rm Pr\,Ta}{\rm Ra}\right)\left(\frac{g}{\Omega^2 r}\right)\partial_\theta \ln T
	\label{eq:xi_temp}
\end{align}
\citet{2012ApJ...756..169A} provide $g$, $R$, $\rm Pr$, $\rm Ta$ and $\rm Ra$.
They also provide a quantity $\Omega/\Omega_\odot$, which we take to be the mean angular velocity of the simulation normalized by the typical solar angular velocity of $4.3\times 10^{-7}{\rm cycles\,s^{-1}}$.
We then estimate
\begin{align}
	\partial_\theta \ln T \approx \frac{3}{\pi} \frac{\Delta T}{T_{\rm base\ CZ}},
\end{align}
where we use the temperature $T_{\rm base\ CZ}$ at the base of the convection zone because the temperature variation is dominated by that region.

We are also interested in the convection speed normalized by $\varv_{\rm c} / h |N|$.
To obtain this we write
\begin{align}
	\frac{\varv_{\rm c}}{h |N|} &\approx \frac{\varv_{\rm c}}{h}\sqrt{\frac{\gamma}{(\gamma-1)|g\partial_r s|}}\\
	&\approx \frac{\varv_{\rm c}}{h}\sqrt{\frac{\gamma r}{(\gamma-1)g |r \partial_r s|}}\\
	&\approx \frac{\varv_{\rm c}}{2 \Omega h}\sqrt{\frac{\gamma}{\left(\frac{\rm Ra}{\rm Pr\,Ta}\right)}}\\
	&\approx \mathrm{Ro} \sqrt{\gamma\frac{\rm Pr\,Ta}{\rm Ra}}.
	\label{eq:vc_hN_Rayleigh}
\end{align}
Finally, we obtain the P\'{e}clet number by equation~\eqref{eq:Pe}.

\subsubsection{Matt 2011 Simulations}
\label{appen:matt}

These data come from three-dimensional anelastic hydrodynamic simulations of convection in a rotating spherical shell domain~\citep{2011AN....332..897M}.
The differential rotation is reported as the difference in angular velocity between the equator and a latitude of $\pi/3$, normalized to the equatorial angular velocity, so we use equation~\eqref{eq:0_60} to compute $|R\nabla\ln\Omega|$.
The Rossby number actually realised in the flow is also reported, so we follow the procedure we used in appendix~\ref{appen:browning04} to calculate $\Omega/|N|$.

The temperature difference between the equator and a latitude of $\pi/3$ is also reported.
We follow the same procedure as in appendix~\ref{appen:aug12} to compute $\xi$ from this, except that we compute
\begin{align}
	g = \frac{G M}{r^2},
\end{align}
where $G$ is Newton's constant and $M$ is the mass of the star because~\citet{2011AN....332..897M} do not report $g$ but do report $M$ and $R$.

We compute the ratio $\varv_{\rm c} / h|N|$ with the Rossby, Prandtl, Taylor and Rayleigh numbers as in appendix~\ref{appen:aug12}.
Finally, we obtain the P\'{e}clet number from equation~\eqref{eq:Pe}.

\subsubsection{Brown 2008 Simulations}
\label{appen:brown08}

These data come from simulations of three-dimensional anelastic hydrodynamic convection in rotating spherical shells, reported in tables~1 and~2 of~\citet{2008ApJ...689.1354B}.
The differential rotation is reported as the kinetic energy of differential rotation, defined to be the mass-weighted average of the squared difference between the rotation rate and the mean rotation rate.
That is, they report
\begin{align}
	{\rm DRKE} \approx \rho R^2 \langle\left( \Omega - \overline{\Omega} \right)^2\rangle,
\end{align}
where $\langle ... \rangle$ denotes the mass-weighted average and $\bar{\Omega}$ is the mean rotation rate.
They further report the convective kinetic energy
\begin{align}
	{\rm CKE} \approx \rho \varv_{\rm c}^2.
\end{align}
To ensure consistent normalization we estimate $\varv_{\rm c}$ from their reported Rossby number as
\begin{align}
	\varv_{\rm c} \approx 2 \Omega \mathrm{Ro},
\end{align}
so that the shear is
\begin{align}
	|R\nabla\Omega| \approx \frac{\varv_{\rm c}}{R}\sqrt{\frac{\rm DRKE}{\rm CKE}} \approx 2\Omega \mathrm{Ro}\sqrt{\frac{\rm DRKE}{\rm CKE}},
	\label{eq:drke}
\end{align}
where $\mathrm{Ro}$ is the Rossby number actually realised in the flow and of the form of equation~\eqref{eq:vortRo}.
We calculate $\Omega/|N|$ from $\mathrm{Ro}$ as in appendix~\ref{appen:browning04}

We compute the ratio $\varv_{\rm c} / h|N|$ with the Rossby, Prandtl, Taylor and Rayleigh numbers as in appendix~\ref{appen:aug12}.
The one difference is that the Rayleigh number is reported as in equation~\eqref{eq:Rayleigh_Nrho}.
So we must divide it by $(\gamma-1)/\gamma$ before using equation~\eqref{eq:vc_hN_Rayleigh}.
As before, we obtain the P\'{e}clet number from equation~\eqref{eq:Pe}.

\subsubsection{Brun 2009 Simulations}
\label{appen:brun09}

These data come from three-dimensional anelastic hydrodynamic simulations of convection in a rotating spherical domain reported by~\citet{2009ApJ...702.1078B}.
The differential rotation is reported as the mass-weighted average of the squared difference between the rotation rate and the mean rotation rate.
We translate this into a shear as we did in appendix~\ref{appen:brown08}.
The Rossby number actually realised in the flow is reported in the form of equation~\eqref{eq:vortRo} and we compute $\Omega/|N|$ from this as in appendix~\ref{appen:browning04}.

We compute the ratio $\varv_{\rm c} / h|N|$ in the same way as we did for the data from~\citet{2012ApJ...756..169A}, using the Rossby, Prandtl, Taylor and Rayleigh numbers.
Two differences are that~\citet{2009ApJ...702.1078B} used a Prandtl number of $1$ and that they report the Rayleigh number as in equation~\eqref{eq:Rayleigh_Nrho}.
So we must divide $\mathrm{Ra}$ by $(\gamma-1)/\gamma$ before using equation~\eqref{eq:vc_hN_Rayleigh}.
They also do not report $\gamma$ so we assume that it is $5/3$.
As before, we obtain the P\'{e}clet number from equation~\eqref{eq:Pe}.

\subsubsection{Brun 2017 Simulations}

These data come from three-dimensional anelastic hydrodynamic simulations of convection in a rotating spherical domain~\citep{2017ApJ...836..192B}.
The differential rotation is reported as the mass-weighted average of the squared difference between the rotation rate and the mean rotation rate.
We translate this into a shear as we did in appendix~\ref{appen:brown08}.
The Rossby number actually realised in the flow is reported in the form of equation~\eqref{eq:vortRo} and we compute $\Omega/|N|$ from this as in appendix~\ref{appen:browning04}.

The temperature difference between the equator and a latitude of $\pi/3$ is also reported.
We follow the same procedure as in appendix~\ref{appen:aug12} to compute $\xi$ from this, except that we compute $g$ as in appendix~\ref{appen:matt}.

We compute the ratio $\varv_{\rm c} / h|N|$ in the same way as we did in appendix~\ref{appen:brun09}.
As before, we obtain the P\'{e}clet number from equation~\eqref{eq:Pe}.

\subsubsection{Featherstone 2015}
\label{appen:featherstone}

These data come from three-dimensional anelastic hydrodynamic simulations of convection in rotating spherical shells, reported by~\citet{Featherstone_2015}.
The differential rotation is reported as the mass-weighted average of the squared difference between the rotation rate and the mean rotation rate.
We translate this into a shear as we did in appendix~\ref{appen:brown08}.
The Rossby number actually realised in the flow is reported in the form of equation~\eqref{eq:vortRo}, and we compute $\Omega/|N|$ from this as in appendix~\ref{appen:browning04}.

\citet{Featherstone_2015} report a modified Rayleigh number
\begin{align}
	\mathcal{R}^* \equiv \frac{g |\partial_r S|}{c_{P} \Omega^2},
	\label{eq:Rstar}
\end{align}
where these quantities are all as defined previously.
Using equations~\eqref{eq:s},~\eqref{eq:N2} and~\eqref{eq:s_dim} we find
\begin{align}
	\mathcal{R}^* = \frac{|N|^2}{\Omega^2},
\end{align}
from which we compute $\Omega/|N|$.

To compute $\varv_{\rm c} / h |N|$ we use the Rossby number actually realised in the flow.
With this we find
\begin{align}
	\frac{\varv_{\rm c}}{h |N|} \approx 2 \mathrm{Ro} \sqrt{\mathcal{R}^*}.
\end{align}

We compute the Reynolds number as $\mathrm{Re} = 2 \Omega L \mathrm{Ro} / \nu$ from the given $\Omega$, $L$ and viscosity $\nu$.
We then compute the P\'{e}clet number from equation~\eqref{eq:Pe}.

\subsubsection{Augustson 2013 \& 2016 Simulations}

These data come from anelastic three-dimensional anelastic hydrodynamical simulations of rotating convection in a spherical domain~\citep{2016ApJ...829...92A} and spherical shells~\citep{0004-637X-777-2-153}.
The differential rotation is reported as the mass-weighted average of the squared difference between the rotation rate and the mean rotation rate.
We translate this into a shear as we did in appendix~\ref{appen:brown08}.
The Rossby number actually realised in the flow is reported in the form of equation~\eqref{eq:vortRo} and we compute $\Omega/|N|$ from this as in appendix~\ref{appen:browning04}.

The temperature difference between the equator and a latitude of $\pi/3$ is also reported by~\citet{0004-637X-777-2-153}.
We follow the same procedure as in appendix~\ref{appen:aug12} to compute $\xi$ from this.
We likewise compute the ratio $\varv_{\rm c} / h|N|$ in the same way as we did in appendix~\ref{appen:aug12}.
We obtain the P\'{e}clet number from equation~\eqref{eq:Pe}.

\subsubsection{Guerrero Simulations}

These data come from simulations of three-dimensional anelastic convection in rotating spherical shells, reported in table~1 of~\citet{2013ApJ...779..176G}.
The reported differential rotation rate is the surface equator-pole difference in $\Omega$ normalised to that of the equator.
Because this spans $\pi/2$ radians we divide by $\pi/2$ to give it the same units as $\partial_\theta\Omega/\Omega$.
We take this as to be approximately $|R\nabla\Omega|/\Omega$.

They report $\mathcal{R}^*$ as in equation~\eqref{eq:Rstar}.
We compute $\Omega/|N|$ from this as in appendix~\ref{appen:featherstone}.

They also report the Rossby number in the form
\begin{align}
	\mathrm{Ro} = \frac{\tilde{v}}{2\bar{\Omega} L},
\end{align}
where $\tilde{v}$ is the root-mean square of the fluctuating velocity, $L$ is the vertical size of the domain and $\bar{\Omega}$ is the rotation rate of the frame with zero total angular momentum.
We use this to find
\begin{align}
	\frac{\varv_{\rm c}}{h |N|} \approx 2 \mathrm{Ro} \sqrt{\mathcal{R}^*}.
\end{align}

We were unable to determine the P\'{e}clet number in these simulations because the EULAG software instrument they employ does not use explicit diffusivities and so makes it difficult to determine precise values for dimensionless parameters such as $\mathrm{Pe}$.

\subsubsection{Mabuchi Simulations (HD)}
\label{appen:mabuchi}

These data come from three-dimensional fully compressible hydrodynamic simulations of rotating convection in spherical shells, reported by~\citet{2015ApJ...806...10M}.
Data were taken from their table 1.
The reported differential rotation is in the form
\begin{align}
	\alpha \equiv \frac{\Omega-\bar{\Omega}}{\bar{\Omega}},
	\label{eq:alpha}
\end{align}
where $\Omega$ is evaluated at the surface of the star on the equator.

The convective Rossby number is reported as
\begin{align}
	\mathrm{Ro}_{\rm conv}^2 \equiv \frac{\mathrm{Ra}}{\mathrm{Ta}\, \mathrm{Pr}},
\end{align}
where
\begin{align}
	\mathrm{Ta} \equiv \left(\frac{2\bar{\Omega} d}{\nu}\right)^2,
	\label{eq:Ta_mab}
\end{align}
$d$ is the vertical size of the domain and
\begin{align}
	\mathrm{Ra} \equiv \frac{g d^4}{\nu \xi}\left(-\frac{\partial_r s}{c_{P}}\right).
	\label{eq:Ra_mab}
\end{align}
With these definitions, the convective Rossby number simplifies to
\begin{align}
	\mathrm{Ro}_{\rm conv}^2 = \frac{|N|^2}{4\Omega^2},
\end{align}
from which we compute $\Omega/|N|$.

The Rossby number actually realised in the flow is also reported as
\begin{align}
	\mathrm{Ro} = \frac{\pi \varv_c}{\Omega d},
	\label{eq:RoMab}
\end{align}
where $d$ is the radial extent of the convection zone.
From this we obtain
\begin{align}
	\frac{\varv_{\rm c}}{h |N|} = \left(\frac{2}{\pi}\right) \left(\frac{\mathrm{Ro}}{\mathrm{Ro}_{\rm conv}}\right).
\end{align}

We obtain the P\'{e}clet number from equation~\eqref{eq:Pe}.
We calculate the Reynolds number using the height of the unstable layer ($0.3$), the thermal diffusivity ($3.9\times 10^{-4}$) and the reported root-mean-square velocities, all in code units.

\subsubsection{K{\"a}pyl{\"a} Simulations}

These data come from simulations of three-dimensional fully compressible hydrodynamical convection in rotating spherical shells~\citep{2011A&A...531A.162K}.
Data were taken from their table~1.
The reported differential rotation rate is the surface difference of $\Omega$ over $2\pi /3$ radians normalised to that of the equator.
We thus divide by $2\pi /3$ to estimate $\partial_\theta\Omega/\Omega$ and take this as to be approximately $|R\nabla\Omega|/\Omega$.

The Rayleigh number is reported as in equation~\eqref{eq:Ra_mab}.
The Coriolis number was reported as $\mathrm{Ro}^{-1}$, with $\mathrm{Ro}$ defined by equation~\eqref{eq:RoMab}.
The Prandtl number is reported as well.
Finally, the Reynolds number is reported as
\begin{align}
	\mathrm{Re} = \frac{\tilde{v} d}{2\pi \nu},
\end{align}
where $d$ is the vertical size of the domain and $\tilde{\nu}$ is the viscosity.

With these pieces we can write the Taylor number, defined by equation~\eqref{eq:Ta_mab}, as
\begin{align}
	\mathrm{Ta} = (\mathrm{4\pi^2 Co}\mathrm{Re})^2
\end{align}
and find
\begin{align}
	\mathrm{Ro}_{\rm conv}^2 = \frac{\mathrm{Ra}}{\mathrm{Ta}\, \mathrm{Pr}} = \frac{|N|^2}{4\Omega^2},
\end{align}
from which we compute $\Omega/|N|$.
We further obtain
\begin{align}
	\frac{\varv_{\rm c}}{h |N|} = \left(\frac{2}{\pi}\right) \left(\frac{\mathrm{Ro}}{\mathrm{Ro}_{\rm conv}}\right).
\end{align}
We calculate the P\'{e}clet number from equation~\eqref{eq:Pe}.

\subsubsection{Gastine 2012 Simulations}
\label{appen:gastine12}

These data come from three-dimensional hydrodynamic simulations of anelastic convection in rotating spherical shells, reported by~\citet{2012Icar..219..428G}.

The polytropic index is reported as $m=2$, so that $\gamma = 3/2$.
The same Ekman number was used in each of these simulations, and reported as
\begin{align}
	\mathrm{Ek} = \frac{\nu}{\Omega_{\rm frame} d^2} = 10^{-4},
\end{align}
where $\Omega_{\rm frame}$ is the angular velocity of the simulation reference frame, $d$ is the vertical size of the domain and $\nu$ is the microscopic viscosity.

The Prandtl number was also held constant and reported as
\begin{align}
	\mathrm{Pr} \equiv \frac{\nu}{\alpha} = 1,
\end{align}
where $\alpha$ is the microscopic thermal diffusivity.
Finally, the aspect ratio $\eta$ was held constant and reported as
\begin{align}
	\eta = \frac{r_{\rm inner}}{r_{\rm outer}}.
\end{align}
This is related to the vertical size $d$ of the domain by
\begin{align}
	\eta = 1 - \frac{d}{r_{\rm outer}}.
\end{align}

The Rayleigh number is reported as
\begin{align}
	\mathrm{Ra} \equiv \frac{g d^3 \Delta S}{c_{p} \nu \alpha},
\end{align}
where $c_{p}$ is the specific heat at constant pressure, $g$ is the acceleration due to gravity and $\Delta S$ is the difference in specific dimensionful entropy across the domain.
The ratio $\Delta S / c_{p}$ may be written as
\begin{align}
	\frac{\Delta S}{c_{p}} = \frac{1}{\gamma} \ln p - \ln \rho = \frac{\gamma-1}{\gamma}\Delta s,
\end{align}
where $s$ is the dimensionless entropy defined in equation~\eqref{eq:s}.
So
\begin{align}
\mathrm{Ra} = \frac{(\gamma-1) g d^3 \Delta s}{\nu \alpha \gamma}.
\label{eq:Ra}
\end{align}
Approximating the entropy gradient as radial, we write
\begin{align}
	g\Delta s \approx g d \frac{\partial s}{\partial r} \approx \gamma d |N|^2,
\end{align}
so that
\begin{align}
\mathrm{Ra} \approx \frac{(\gamma-1)d^4 |N|^2}{\nu \alpha}
\label{eq:RaGastine}
\end{align}
and, with these pieces, we may write the ratio
\begin{align}
	\frac{\Omega}{|N|} \approx \sqrt{\frac{(\gamma-1)\mathrm{Pr}}{\mathrm{Ek}^2 \mathrm{Ra}}}.
	\label{eq:OmegaNfromRa}
\end{align}

The ratio of the axisymmetric azimuthal kinetic energy in the frame rotating at $\Omega_{\rm frame}$ to the total kinetic energy was reported.
In the language of~\citet{2008ApJ...689.1354B} this ratio is
\begin{align}
	\frac{\rm{DRKE}}{\rm{DRKE} + \rm{CKE}},
\end{align}
from which we can compute $\rm{DRKE}/\rm{CKE}$.

The Rossby number was also reported as
\begin{align}
	\mathrm{Ro} \equiv \mathrm{Ek} \mathrm{Re} = \frac{\nu}{\Omega d^2}\left(\frac{\langle \varv^2\rangle^{1/2}}{\nu d}\right) = \frac{\langle \varv^2\rangle^{1/2}}{\Omega d},
\end{align}
where we have inferred the definition of $\mathrm{Re}$ from the text surrounding their equation (26) and $\langle \varv^2\rangle^{1/2}$ is the time- and volume-averaged root-mean square of the velocity.
Putting these two numbers together we find
\begin{align}
	|R\nabla\ln\Omega| &\approx \frac{\langle \varv^2\rangle^{1/2}}{\Omega R} \sqrt{\frac{\mathrm{DRKE}}{\mathrm{CKE}}}\\
	&\approx \frac{d}{r_{\rm outer}\mathrm{Ro}} \sqrt{\frac{\mathrm{DRKE}}{\mathrm{CKE}}}\\
	&\approx \left(1-\eta\right) \sqrt{\frac{\mathrm{DRKE}}{\mathrm{CKE}}}.
\end{align}

We computed the ratio of the convection speed to $h|N|$ as
\begin{align}
	\frac{\varv_{\rm c}}{h |N|} \approx \frac{\varv_{\rm c} n }{d |N|} = n\left(\frac{\Omega}{|N|}\right)\left(\frac{\varv_c}{\Omega d}\right) = n \frac{\Omega}{|N|}\mathrm{Ro} \sqrt{1-\frac{\mathrm{DRKE}}{\mathrm{CKE}}},
\end{align}
where the final square root factor corrects the Rossby number for the fraction of energy actually in convection, as opposed to differential rotation, and
\begin{align}
	n \equiv \max\left(1, \frac{N_\rho}{\gamma}\right)
	\label{eq:n}
\end{align}
is the number of pressure scale heights in the domain, which we cut off at unity because that is the relevant mixing length in the limit of no density stratification.
Here $N_\rho$ is the reported number of density scale heights in the domain.
We obtain the P\'{e}clet number using the relation
\begin{align}
	\mathrm{Re} = \mathrm{Ek}^{-1} \mathrm{Ro},
	\label{eq:ReRo}
\end{align}
where $\mathrm{Ek}$ is the Eckman number, and use this Reynolds number in equation~\eqref{eq:Pe}.

\subsubsection{Gastine 2013 Simulations}

These data are from three-dimensional hydrodynamic simulations of anelastic convection in rotating spherical shells, reported by~\citet{2013Icar..225..156G}.
The Prandtl number, aspect ratio, domain size, poltropic indices are the same as those in appendix~\ref{appen:gastine12}.
The Rayleigh number and Eckman number are reported in the same manner but vary from simulation to simulation.
We compute the Rossby number from these in the same way as in appendix~\ref{appen:gastine12}.

\citet{2013Icar..225..156G} report the time-averaged Rossby number at the surface on the equator ($\mathrm{Ro}_{\rm e}$).
We interpret this directly as a measure of differential rotation, so that
\begin{align}
	|R\nabla\ln\Omega| = \mathrm{Ro}_{\rm e} \equiv \frac{\Omega_{\rm equator, surface} - \Omega_{\rm frame}}{\Omega_{\rm frame}},
\end{align}
where $\Omega_{\rm frame}$ is defined as in appendix~\ref{appen:gastine12}.

We compute the ratio of the convection speed to $h|N|$ with equation~\eqref{eq:RaGastine} to be
\begin{align}
	\frac{\varv_{\rm c}}{h |N|} \approx \frac{\varv_{\rm c} n }{d |N|} \approx n \sqrt{\frac{\rm{Re}^2 (\gamma-1)\rm{Pr}}{\mathrm{Ek}^2 \rm{Ra}}},
\end{align}
where $n$ is defined by equation~\eqref{eq:n} and $\rm{Re}$ is the Reynolds number of fluctuations in the flow.

We obtain the P\'{e}clet number from equation~\eqref{eq:Pe}.
Here we use the fluctuating Reynolds number, because the other Reynolds numbers reported by~\citet{2012Icar..219..428G} are broken down into specific spatial components.

\subsubsection{Aurnou Simulations}
\label{appen:aurnou}

These data are from simulations of three-dimensional hydrodynamic Boussinesq convection in rotating spherical shells~\citep{2007Icar..190..110A}.
They appear in the second summary figure of~\citet{2014MNRAS.438L..76G} and were extracted with automated graphic data extraction software to provide $\alpha$ as in equation~\eqref{eq:alpha}.
Points with $\alpha < 0.02$ were removed from the sample because the plot was scaled linearly and below this limit the figure resolution was insufficient to ensure accuracy.
To measure the rotation we note that $\alpha$ is provided as a function of the convective Rossby number, which~\citet{2014MNRAS.438L..76G} define
\begin{align}
	\mathrm{Ro}_{\rm c} = \sqrt{\frac{\mathrm{Ek}^2\mathrm{Ra}}{\mathrm{Pr}}}.
\end{align}
Inspection of equations~(2-4) of~\citep{2007Icar..190..110A} reveals that this quantity is just $|N|/\Omega$, from which we compute $\Omega/|N|$.

We obtain the P\'{e}clet number from equations~\eqref{eq:Pe} and~\eqref{eq:ReRo}.
To find the Ekman number we note that the quantity $\mathrm{Ra}_{*}$ reported by~\citet{2014MNRAS.438L..76G} is not just the convective Rossby number, but rather
\begin{align}
	\mathrm{Ra}_* = \mathrm{Ra} \frac{\mathrm{Pr}}{\mathrm{Ek}^2}.
\end{align}
From this we can solve for $\mathrm{Ek}$ and thence for $\mathrm{Pe}$.

\subsubsection{Kaspi Simulations}

These data are from three-dimensional anelastic general circulation simulations over a thick spherical shell~\citep{2009Icar..202..525K}.
They appear in the second summary figure of~\citet{2014MNRAS.438L..76G} and were extracted with automated graphic data extraction software providing the quantity $\alpha$ given in equation~\eqref{eq:alpha}.
These were extracted and processed as in appendix~\ref{appen:aurnou}.

\subsubsection{Gilman Simulations}

These data come from simulations of three-dimensional Boussinesq convection in rotating spherical shells reported in~\citet{1977GApFD...8...93G} and~\citet{1979ApJ...231..284G}.
The results appear in the second summary figure in~\citet{2014MNRAS.438L..76G}.
These were extracted and processed as in appendix~\ref{appen:aurnou}.

\subsubsection{Rogers Simulations}

These data are from three-dimensional hydrodynamic simulations of anelastic convection in a spherical domain reported by~\citet{2015ApJ...815L..30R}.
The results were taken from their table.
The table lists $\varv_c$ for the two extremal convective forcing functions and so we linearly interpolate these as
\begin{align}
	\varv_c = \left[2.9 + (q - 1.5) * (4.5 - 2.9)\right] \mathrm{km\,s^{-1}},
\end{align}
where $q$ equals their $\bar{Q}/\varv_\mathrm{c}$ and is the convective forcing.
We take the scale height to be the size of their convection zone, which is $0.3 R_\odot$.
Using the $\Omega$ values in their table we obtain $\varv_c / h \Omega$.
We then employ the scaling in table~\ref{tab:summary_num} to write
\begin{align}
	\frac{\Omega}{|N|} = \sqrt{\frac{\Omega h}{\varv_c}}
	\label{eq:on_from_vc}
\end{align}
when $\Omega > |N|$ and
\begin{align}
	\frac{\Omega}{|N|} = \frac{\Omega h}{\varv_c}
	\label{eq:on_from_vc}
\end{align}
when $\Omega < |N|$.
Note that this definition assumes that our scaling for the convective velocity holds, and so we cannot use these data to test that scaling.

The differential rotation is reported as the ratio of the core-to-envelope angular velocity.
We estimate the shear from this with equation~\eqref{eq:core_env}.

To obtain the Reynolds number we use their reported viscosity $\nu = 4\times 10^{13}\mathrm{cm^2\,s^{-1}}$ and let
\begin{align}
	\mathrm{Re} = \frac{\varv_c h}{\nu}.
\end{align}
We then determine the P\'{e}clet number through equation~\eqref{eq:Pe} using the Prandtl number
\begin{align}
	\mathrm{Pr} = \frac{\nu}{\alpha},
\end{align}
where the thermal conductivity $\alpha = 5\times 10^{11}\mathrm{cm^2\,s^{-1}}$.

\subsubsection{Raynaud Simulations}

These data are from simulations of three-dimensional anelastic hydrodynamic convection in rotating spherical shells, reported in table~B.1 of~\citet{2018A&A...609A.124R}.
The table contains the number of density scale heights in the simulation domain ($N_\rho$), the ratio of the outer radius of the domain to the inner radius ($\chi$) and the Nusselt number ($\rm Nu$) at both the equator and pole.
The Nusselt number is proportional to the radial entropy gradient.
The latitudinal derivative of this is then
\begin{align}
	\partial_\theta {\rm Nu} \propto \partial_r \partial_\theta s,
\end{align}
so
\begin{align}
	\partial_\theta \ln {\rm Nu} = \frac{\partial_r \partial_\theta s}{\partial_r s}.
\end{align}
If we estimate that the radial derivative in the numerator to produce a factor of order the pressure scale height then
\begin{align}
	\partial_\theta \ln {\rm Nu} \approx \frac{\partial_\theta s}{h \partial_r s}.
\end{align}
By equation~\eqref{eq:xi_approx} we then obtain
\begin{align}
	\xi \approx \frac{h}{r} \partial_\theta \ln {\rm Nu}.
	\label{eq:xi_nu}
\end{align}

The polytropic index $n$ is not specified, so we assume $n=3/2$.
The pressure scale height is then related to the density scale height by a factor of $\gamma = 5/3$, so
\begin{align}
	\frac{h}{r} = \frac{\gamma h_\rho}{r},
\end{align}
where $h_\rho$ is the density scale height.
The average density scale height in the simulation domain is just the size of the domain divided by the number of scale heights it contains, so
\begin{align}
	\frac{h}{r} \approx \frac{\gamma \Delta r}{N_\rho r} = \frac{\gamma}{N_\rho}(1-\chi).
\end{align}
Inserting this into equation~\eqref{eq:xi_nu} we find
\begin{align}
	\xi \approx \frac{\gamma}{N_\rho}(1-\chi)\partial_\theta \ln {\rm Nu}.
	\label{eq:xi_nu2}
\end{align}
We then estimate
\begin{align}
	\partial_\theta \ln {\rm Nu} \approx \frac{2}{\pi} \left|2\frac{{\rm Nu}_{\rm equator} - {\rm Nu}_{\rm pole}}{{\rm Nu}_{\rm equator} + {\rm Nu}_{\rm pole}}\right|
\end{align}
and with equation~\eqref{eq:xi_nu2} obtain $\xi$.

The table contains the Ekman number
\begin{align}
	\mathrm{Ek} = \frac{\nu}{\Omega_{\rm frame} d^2}.
\end{align}
The Rayleigh number is reported as
\begin{align}
\mathrm{Ra} = \frac{g d r^2 \Delta s}{\nu \alpha},
\end{align}
where we have used $\alpha = \kappa c_p$ to convert from their notation to ours.
Using $d/r = 1-\chi$ we see that this Rayleigh number differs from that of equation~\eqref{eq:Ra} by a factor of $\gamma (\gamma-1)^{-1} (1-\chi)^{-2}$, so we may use equation~\eqref{eq:OmegaNfromRa} to find
\begin{align}
	\frac{\Omega}{|N|} \approx \sqrt{\frac{\gamma\mathrm{Pr}}{\mathrm{Ek}^2 (1-\chi)^2 \mathrm{Ra}}}.
\end{align}

%%%---------- close: hydro_sims_appen

%%%%%%%%% jump to MHD_sims_appen
%%%---------- open: MHD_sims_appen

\subsection{MHD Simulations}

We describe how we extract the relevant data from various MHD simulations.

\subsubsection{Soderlund Simulations}

These data are from simulations of three-dimensional Boussinesq MHD convection in rotating spherical shells of varying thickness~\citep{2013Icar..224...97S}.
These appear in the second summary figure in~\citet{2014MNRAS.438L..76G} and were extracted with automated graphic data extraction software, providing the quantity $\alpha$ given in equation~\eqref{eq:alpha}.
Points with $\alpha < 0.02$ were removed from the sample because the plot was scaled linearly and below this limit the figure resolution was insufficient to ensure accuracy.

To measure the rotation we note that $\alpha$ is provided as a function of the convective Rossby number, defined as $|N|/\Omega$.
We obtain the P\'{e}clet number from equations~\eqref{eq:Pe} and~\eqref{eq:ReRo}.

\subsubsection{Mabuchi Simulations (MHD)}

These data are from three-dimensional fully compressible MHD simulations of rotating convection in spherical shells, reported by~\citet{2015ApJ...806...10M}.
Data were taken from their table 1.
The reported differential rotation is in the form of $\alpha$ as defined by equation~\eqref{eq:alpha}.
The data analysis was performed in the same way as for the hydrodynamic case in appendix~\ref{appen:mabuchi}.
\citet{2015ApJ...806...10M} also report the magnetic energy and the kinetic energy in the meridional plane and we take the ratio of these to be $\varv_{\rm A}^2 / \varv_{\rm c}^2$.

We obtain the P\'{e}clet number from equation~\eqref{eq:Pe}.
We calculate the Reynolds number using the height of the unstable layer ($0.3$), the thermal diffusivity ($3.9\times 10^{-4}$), and the reported root-mean-square velocities, all in code units.

\subsubsection{Brun 2005 Simulations}

These data are from three-dimensional anelastic MHD simulations of convection in a rotating spherical domain, reported by~\citet{0004-637X-629-1-461}.
The differential rotation is reported as the mass-weighted average of the squared difference between the rotation rate and the mean rotation rate.
We translate this into a shear as we did for the data of~\citet{2008ApJ...689.1354B}, taking into account the scaling of $\varv_{\rm c}$ in MHD systems given in Table~\ref{tab:summary}.

The Rossby number actually realised in the flow is also reported.
This they define as
\begin{align}
	\mathrm{Ro} = \frac{\varv_c}{2 \Omega R}.
\end{align}
For this calculation we take $h \approx R$ because the average depth is comparable to the radius.
We then calculate $\Omega/|N|$ from this as in appendix~\ref{appen:browning04}.

The magnetic energy and the convective kinetic energy are also reported.
We use this to calculate $\varv_{\rm A}^2 / \varv_{\rm c}^2$.
We compute the ratio $\varv_{\rm c} / h|N|$ in the same way as we did in appendix~\ref{appen:browning04}.
The one difference is that~\citet{0004-637X-629-1-461} do not report the adiabatic exponent $\gamma$, so we assume it is $5/3$.

\citet{2008ApJ...689.1354B} report the P\'{e}clet number so we do not need to compute it.

\subsubsection{Augustson 2013 \& 2016 Simulations}

These data come from anelastic three-dimensional MHD simulations of rotating convection in a spherical domain~\citep{0004-637X-777-2-153} and spherical shells~\citep{2016ApJ...829...92A}.
The differential rotation is reported as the mass-weighted average of the squared difference between the rotation rate and the mean rotation rate.
We translate this into a shear as we did in appendix~\ref{appen:brown08}, taking into account the scaling of $\varv_{\rm c}$ in MHD systems given in Table~\ref{tab:summary}.
The data analysis was performed in the same way as for the hydrodynamic case.
The magnetic energy and the convective kinetic energy are also reported.
We use this to calculate $\varv_{\rm A}^2 / \varv_{\rm c}^2$.
We obtain the P\'{e}clet number from equation~\eqref{eq:Pe}.

\subsubsection{Varela 2016 Simulations}

These data are from anelastic three-dimensional MHD simulations of rotating convection in a spherical shell domain, reported in tables~1 and~2 of~\citet{2016AdSpR..58.1507V}.
The differential rotation is reported as the difference in angular velocity between the equator and a latitude of $\pi/3$, normalized to the equatorial angular velocity, so we follow equation~\eqref{eq:0_60} to compute $|R\nabla\ln\Omega|$.

The Rossby number actually realised in the flow is reported in the form of equation~\eqref{eq:vortRo}.
We calculate $\Omega/|N|$ from this as in appendix~\ref{appen:browning04}.
The magnetic energy and the convective kinetic energy are also reported and used to calculate $\varv_{\rm A}^2 / \varv_{\rm c}^2$.

We compute the ratio $\varv_{\rm c} / h|N|$ in the same way as we did in appendix~\ref{appen:aug12}, using the Rossby, Prandtl, Taylor and Rayleigh numbers.
The one difference is that the Rayleigh number is reported as in equation~\eqref{eq:Rayleigh_Nrho}, so we must divide it by $(\gamma-1)/\gamma$ before using equation~\eqref{eq:vc_hN_Rayleigh}.
We obtain the P\'{e}clet number from equation~\eqref{eq:Pe}.

\subsubsection{Yadav 2013 Simulations}

These data are from Boussinesq three-dimensional MHD simulations of rotation convection in a spherical shell domain, reported in table~A.2 of~\citet{2013Icar..225..185Y}.
The Rayleigh number is reported as
\begin{align}
	\mathrm{Ra} = \frac{g d^3 \Delta T}{\alpha \nu}\frac{d\ln V}{dT},
\end{align}
where $d\ln V/dT$ is the thermal expansivity.
The convention given in their equation~(1) shows that the modified Rayleigh number
\begin{align}
	\mathrm{Ra}^* = \frac{\mathrm{Ra} \mathrm{Ek}^2}{\mathrm{Pr}}
\end{align}
equals $|N|^2/\Omega^2$, so we compute this from the given $\mathrm{Ra}$, $\mathrm{Ek}$ and $\mathrm{Pr}$ and thereby obtain $\Omega/|N|$.

\citet{2013Icar..225..185Y} provide the Rossby number number, from which we obtain the Reynolds number
\begin{align}
	\mathrm{Re} = \mathrm{Ro} \mathrm{Ek}^{-1}.
\end{align}
We then find
\begin{align}
	\frac{\varv_{\rm c}}{h |N|} = \frac{\mathrm{Re} \nu}{d h |N|}.
\end{align}
Letting $h \approx d$ and using $\nu = \Omega d^2 \mathrm{Ek}$ we find
\begin{align}
	\frac{\varv_{\rm c}}{h |N|} = \frac{\mathrm{Re} \mathrm{Ek} \Omega}{|N|} = \mathrm{Re} \mathrm{Ek}\sqrt{\frac{\mathrm{Pr}}{\mathrm{Ra} \mathrm{Ek}^2}} = \mathrm{Re}\sqrt{\frac{\mathrm{Pr}}{\mathrm{Ra}}}.
\end{align}
We use their $\mathrm{Ro}_{\ell}$ in place of $\mathrm{Ro}$ because~\citet{2013Icar..225..185Y} report the former and say that it is more reflective of the ratio of inertial to Coriolis forces than the latter.
The zonal Rossby number is also provided, and we interpret this as giving
\begin{align}
	|R\nabla\ln \Omega| \approx \mathrm{Ro}_{\rm zonal}.
\end{align}

We next compute the P\'{e}clet number as
\begin{align}
	\mathrm{Pe} = \mathrm{Re} \mathrm{Pr}.
\end{align}

Finally, the Lorentz number is reported as
\begin{align}
	\mathrm{Lo} \equiv \langle \frac{B^2}{4\pi \rho \Omega^2 d^2} \rangle^{1/2},
\end{align}
where $\langle ... \rangle$ denotes a volume average and $d$ is the domain thickness.
From this we obtain
\begin{align}
	\varv_{\rm A}^2 = \mathrm{Lo}^2 \Omega^2 d^2,
\end{align}
so
\begin{align}
	\frac{\varv_{\rm A}^2}{\varv_{\rm c}^2} \approx \mathrm{Lo}^2 \mathrm{Ro}_{\ell}^{-2}.
\end{align}
%%%---------- close: MHD_sims_appen

\subsection{Solar Data}
\label{appen:solar}

A helioseismic inversion of the solar profile was obtained from~\citet{Private}.
It corresponds to that appearing in~\citet{2008ApJ...681..680A}.
This includes $\Omega$ as a function of position throughout the convective envelope as well as in the upper portions of the radiative envelope.
This profile was then supersampled on to a grid running from $0.5 R_\odot$ to $R_\odot$ in the radial direction, with $100$ uniformly spaced points, and from $0$ to $\pi$ in the latitude, with $70$ uniformly spaced points.
By applying a differentiating Gaussian filter with width equal to five grid points in each of the radial and latitudinal directions we computed $R\nabla\Omega/\Omega$ from this profile.
We then averaged each component of this over latitude, weighted by $\sin\theta$, and took the square root of the result to produce a measure of the mean radial and latitudinal differential rotation at each radial slice.

The radial profile of $|N|$ was obtained from a \code{MESA} model of a $1\,M_\odot$ star with metallicity $Z=0.02$ at an age of $4.6\,\mathrm{Gyr}$.
This was used to compute $\Omega/|N|$ everywhere in the solar convection zone.
We then averaged $\Omega^2/|N|^2$ over latitudes, weighting by $\sin\theta$, and computed the square root to determine the mean $\Omega/|N|$ in each radial slice.

\citet{1972SoPh...23..257A} provide an upper bound on the temperature difference between the solar pole and equator.
Following equation~\eqref{eq:xi_approx} we write
\begin{align}
	\xi \approx \frac{\partial_\theta s}{r \partial_r s}.
\end{align}
We then note that, from equation~\eqref{eq:N2},
\begin{align}
	\partial_r s \approx \frac{\gamma}{\gamma-1} \frac{|N|^2}{g}
\end{align}
so
\begin{align}
	\xi \approx \frac{g (\gamma-1)}{\gamma r |N|^2} \partial_\theta s
\end{align}
and inserting equation~\eqref{eq:dsdlnT} we find
\begin{align}
	\xi \approx \frac{g}{r |N|^2} \partial_\theta \ln T.
	\label{eq:xi_solar}
\end{align}
Because the temperature measurement is at the surface we compute $\xi$ using $|N|^2$, $g$ and $r$ at the surface, estimating
\begin{align}
	\partial_\theta \ln T \approx \left(\frac{2}{\pi}\right)\frac{T_{\rm pole}-T_{\rm equator}}{T_{\rm eff}}.
\end{align}
The factor of $2/\pi$ accounts for the number of radians involved in estimating the latitudinal derivative from a finite difference.
To plot this we take the rotation rate of the Sun to be its mean rate and compute $\Omega/|N|$ for the surface $|N|$.

\subsection{Jupiter Data}
\label{appen:Jupiter}

The profile of differential rotation in the surface layers of Jupiter was taken to be that given by~\citet{2018Natur.555..223K}.
This yields the variation of the characteristic zonal flow rate with depth but does not provide the shear itself.
To compute this we first note that roughly 30 per-cent of Jupiter's surface is covered in latitudinal bands with velocities of order $100\,\mathrm{km\,s^{-1}}$, while the remaining latitudes contain bands with velocities of order $25\mathrm{km\,s^{-1}}$~\citep{2018Natur.555..223K}.
The former are also approximately twice as wide per band as the latter.
There are approximately $15$ bands in total, and so we estimate the typical shear at the surface to be
\begin{align}
	|R\nabla\Omega| \approx \frac{15}{\pi R_\mathrm{J}} \left(0.3 \times 25\mathrm{km\,s^{-1}} + \frac{1}{2}\times 0.7 \times 100\mathrm{km\,s^{-1}}\right),
\end{align}
where $R_\mathrm{J} \approx 7\times 10^{9}\mathrm{cm}$ is the radius of Jupiter.
We then assume that the band velocities scale with depth following the profile
\begin{align}
	\varv_{\phi,\,\mathrm{band}} \propto \left[(1-\alpha)e^{(r-\alpha)/H}+\alpha\frac{1+\tanh\left(\frac{r-a-H}{\Delta H}\right)}{1+\tanh\left(\frac{H}{\Delta H}\right)}\right]
\end{align}
found by~\citet{2018Natur.555..223K}, where $\alpha$, $H$, $a$ and $\Delta H$ are given therein.
The shear then scales as
\begin{align}
	|R\nabla \Omega| \propto \frac{\varv_{\phi,\,\mathrm{band}}}{r}.
\end{align}
We further assume that the mean rotation period remains $9.92\mathrm{hr}$~\citep{2018Natur.555..223K}.

For regions deeper than $3000\,\mathrm{km}$ we bound the band velocity above by the inferred $6\,\mathrm{m\,s^{-1}}$~\citep{2018Natur.555..227G} and use the band structure above to compute the shear.
Furthermore we treat the number of bands as varying linearly in radial coordinate between $1$ near the centre of the planet and $15$ at the surface.
This is an approximation of the cylindrical nature that~\citet{2018Natur.555..223K} assumed for Jupiter's differential rotation.
Following this assumption we consider this estimate to be for both the spherical-radial and latitudinal shears.

All that remains is to compute $|N|_0$.
To do this we ran a \code{MESA} model of a Jupiter-mass planet irradiated by $12500\,{\rm erg\,cm^{-2}\,s^{-1}}$ at a pressure of $1\,{\rm bar}$.
This produced a model within $3$~per~cent of Jupiter's radius and an effective temperature of $136\,{\rm K}$, close to the Jovian $1\,{\rm bar}$ temperature of $165\,{\rm K}$~\citep{2004jpsm.book..105I}.
To ensure efficient convection we exclude the photosphere and the top scale height of the convection zone from our analysis.
The resulting profile of $|N|_0$ is quite close to what the order of magnitude estimate
\begin{align}
	|N|_0 \approx \left(\frac{F}{\rho z^3}\right)^{1/3}
\end{align}
would suggest.
Here $F$ is the heat flux, $\rho$ is the density and $z$ is depth from the surface.

%%%---------- close: MHD_sims_appen
%%%%%%%%%%% jump to expansion
%%%---------- open: expansion
\section{Expansion}

\label{appen:expansion}
Using equations~\eqref{eq:s} and~\eqref{eq:N2} we may relate the increase in $|N|$ to a change in the pressure and density gradients.
For a radial gravity field
\begin{align}
	N^2 = -\frac{1}{\gamma} g \left(\partial_r \ln P - \gamma \partial_r \ln \rho\right).
\end{align}
Inserting equation~\eqref{eq:h} and rearranging, we find
\begin{align}
	\frac{d\ln \rho}{d\ln p} \approx \frac{1}{\gamma}-\frac{h N^2}{g}.
\end{align}
For $\Omega \gg |N|_0$ in the MHD limit the density gradient changes by an amount
\begin{align}
	\delta\left(\frac{d\ln \rho}{d\ln p}\right) \approx \frac{h}{g}(|N|_0^2-|N|^2) \approx -\frac{h}{g}|N|^2 \approx -\frac{\Omega^{2/7}}{|N|_0^{2/7}}\left(\frac{h N_0^2}{g}\right).
\end{align}
If this change in the density gradient results in a change in the mean density of the convection zone of the same order, the radius of the star is changed by an amount
\begin{align}
	\frac{\delta R}{R} \approx -\frac{d_{\mathrm{cz}}}{R}\delta\left(\frac{d\ln \rho}{d\ln p}\right) \approx \frac{d_{\mathrm{cz}}}{R}\frac{\Omega^{2/7}}{|N|_0^{2/7}}\left(\frac{h N_0^2}{g}\right),
\end{align}
where $d_{\mathrm{cz}}$ is the depth of the convection zone.

By comparison the bloating owing to centrifugal effects is
\begin{align}
	\frac{\delta R}{R} \approx \frac{\Omega^2 R}{g},
\end{align}
which is larger by a factor
\begin{align}
	\frac{(\delta R/R)_{\mathrm{centrifugal}}}{(\delta R/R)_{\mathrm{convective}}} \approx \frac{R^2}{h d_{\mathrm{cz}}}\left(\frac{\Omega}{|N|_0}\right)^{12/7} \gg 1.
\end{align}
%%%---------- close: expansion
%%%%%%%%%%% jump to Navg
%%%---------- open: Navg
\section{Sensitivity to $|N|$ averaging}
\label{appen:Navg}

\begin{figure}
\centering
\includegraphics[width=0.47\textwidth]{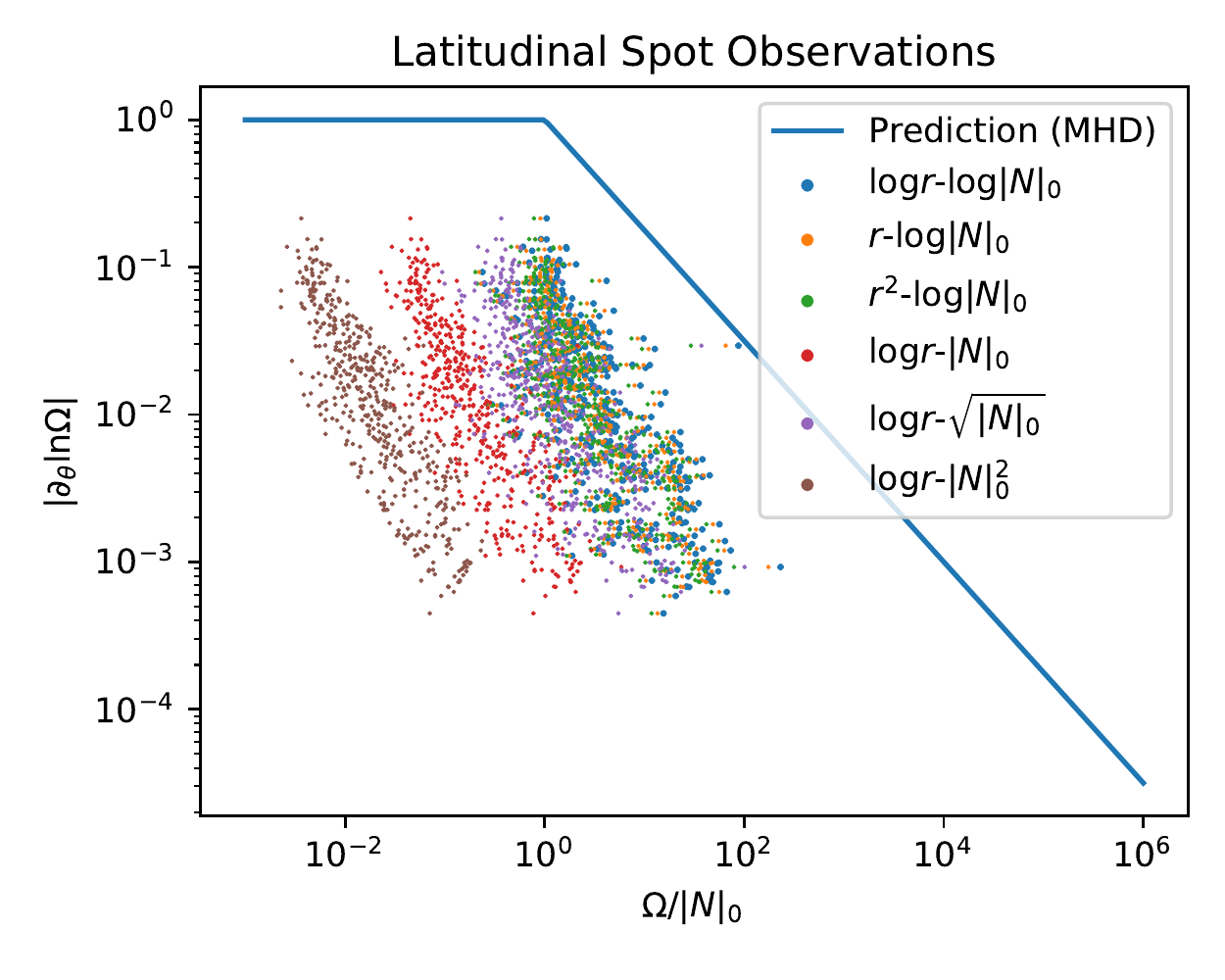}
\caption{The relative latitudinal shear $|\partial_\theta \ln\Omega|$ is shown as a function of $\Omega/|N|_0$ for observed convecting stars inferred from star spot measurements by~\citet{2017AJ....154..250L} along with our predictions. The same data are shown with different colours representing different averaging schemes for $|N|_0$ which weight by different factors in radius and average different functions of $|N|_0$.}
\label{fig:Navg}
\end{figure}

As described in Appendix~\ref{appen:data} and Section~\ref{sec:latitude}, stellar models produce profiles of $|N|_0$ rather than a single number.
To compute an average $|N|_0$ from such a model for comparison, we have had to make a choice (equation~\eqref{eq:N_avg}) of how to weight different parts of the stellar model.
In Fig.~\ref{fig:Navg} we show the results of a variety of different choices on the shape of the latitudinal shear data from~\citet{2017AJ....154..250L}.
In particular, we show weighting by $\log r$, $r$, and $r^2$ and averaging in $\log |N|_0$, as well as weighting by $\log r$ and averaging in $\log |N|_0$, $|N|_0$, $\sqrt{|N|_0}$, and $|N|_0^2$.
Different choices of averaging and weighting produce very different overall scales for $|N|_0$, by up to a factor of $10^2$, but do a good job at preserving the shape of the data (i.e. slope of shear versus $\Omega$).
Because we are primarily comparing our theory with the shape of the data, and admit uncertainty as to the transition between the slowly-rotating and rapidly-rotating limits, the precise choice of averaging therefore makes only a small difference to our comparisons.

\section{Sensitivty to stellar modelling}
\label{appen:models}

In addition to the systematic uncertainty in $|N|_0$ arising from uncertainties in averaging and weighting, there are uncertainties involved in matching stellar models to the observed characteristics of stars.
In matching main-sequence models to observations we have relied on the observed effective temperatures, and for giants with asteroseismic observations we have directly used the inferred masses, ages and metallicities.

These matching exercises are subject to several uncertainties.
First, different stellar models constructed with different software instruments typically disagree on observable properties at the $1-5$~per-cent level even with identical input physics~\citep{2015A&A...575A.117S}.
Next, input physics like opacities are uncertain at the $10-20$~per-cent level~\citep{2015Natur.517...56B}.
In addition different choices of age on the main-sequence, or of Helium abundance, can also lead to changes in effective temperature on this level.

Most importantly, though, 1D stellar models rely on mixing length theory, which is a highly simplified prescription for convection.
One known source of uncertainty in this theory is the mixing length parameter $\alpha_{\mathrm{MLT}}$, which is seen to vary by as much as $20$~per-cent between stars which differ by $0.2 M_\odot$~\citep{2018ApJ...856...10J}.
Naively extrapolating this over the range of stellar masses we have considered, this perhaps leads to a factor of $2$ uncertainty in $\alpha_{\mathrm{MLT}}$.
In limit of efficient convection, $|N|_0 \propto \alpha_{\mathrm{MLT}}^{-2/3}$~\citep[see e.g.][and note that $|N|_0 \propto \sqrt{\nabla-\nabla_{\rm ad}} \propto \sqrt{\nabla-\nabla'}$]{2019ApJ...883..106C}, so the dominant source of uncertainty in $|N|_0$ would seem to be that in the mixing length, and more broadly in using such a simplified theory for convection in the first place.%%%---------- close: Navg

\bsp	% typesetting comment
\label{lastpage}
\end{document}